%% file: block.tex
\documentclass[12pt]{article}
\usepackage{amsmath,amstext, amssymb,graphicx,fancybox,subfigure,amsthm}
\newtheorem{theorem}{Theorem}

\theoremstyle{definition}

\newtheorem{notation}[theorem]{Notation}

\theoremstyle{remark}

\def\cG{\mathcal{G}}
\def\EE{\mathcal{E}}
\def\cA{\mathcal{A}}

\def\cP{\mathcal{P}}

\def\cM{\mathcal{M}}

\def\HH{\mathcal{H}}
\def\KK{\mathcal{K}}
\def\CC{{\Bbb C}}
\def\NN{{\Bbb N}}

\def\ff{\varphi}

\def\entspricht{\qquad\hat =\qquad}

\def\tr{\text{\rm{tr}}}

\def\ee{\varepsilon}

\def\kk{\kappa}                
\def\Tr{\text{\rm{Tr}}}   

\def\cH{\HH}

\newcommand{\comment}[1]{}

\title{Spectra of large block matrices}

\author{ R. Rashidi Far \and T. Oraby \and W.
Bryc\thanks{Research partially supported by NSF grant
\#DMS-0504198.}\and R. Speicher\thanks{Research supported by a Discovery Grant
and a Leadership Support Initiative Award from the Natural Sciences and Engineering
Research Council of Canada and by a Killam Fellowship from the Canada Council for
the Arts}\thanks{R. Rashidi Far and R. Speicher are with the Department of Mathematics and Statistics, Queen's University, Ontario, Canada K7L 3N6 reza, speicher@mast.queensu.ca , T. Oraby and W. Bryc are with the Department of Mathematical Sciences, University of Cincinnati, 28855, Campus Way PO Box 210025, Cincinnati, OH 45221-0025, USA, orabyt@math.uc.edu, wlodzimierz.bryc@uc.edu . } }

\begin{document}
\input{bilder}

\input{bild2}
\date{}
\maketitle

\begin{abstract}
In a frequency selective slow-fading channel in a MIMO system, the channel matrix is of the form of a block matrix. This paper proposes a method to calculate the limit of the eigenvalue distribution of block matrices if the size of the blocks tends to infinity. While it considers random matrices, it takes an operator-valued free probability approach to achieve this goal. Using this method, one derives a system of equations, which can be solved numerically  to compute the desired eigenvalue distribution. \newline The paper initially tackles the problem for square block matrices, then extends the solution to rectangular block matrices. Finally, it deals with Wishart type block matrices. For two special cases, the results of our approach are compared with results from simulations. The first scenario investigates the limit eigenvalue distribution of block Toeplitz matrices. The second scenario deals with the distribution of Wishart type block  matrices for a frequency selective slow-fading channel in a MIMO system for two different cases of $n_R=n_T$ and $n_R=2n_T$. Using this method, one may calculate the capacity and the Signal-to-Interference-and-Noise Ratio in large MIMO systems.\end{abstract}

\section{Introduction}
With the introduction of some sophisticated communication techniques such as CDMA (Code-Division
Multiple-Access) and MIMO (Multiple-Input Mul\-tiple-Output), the communications community has been
looking into analyzing different aspects of these systems, ranging from the channel capacity to the
structure of the receiver. It has been shown that the channel matrix plays a key role in the capacity of the channel \cite{Foschini-Gans-98,Verdu-86} as well as in the structure of the optimum receiver \cite{Madhow-Honig-94,Verdu-98}. More precisely, the eigenvalue distribution of the channel matrix is the factor of interest in different applications. 

\begin{figure}
\centering{
  \includegraphics[width=12cm]{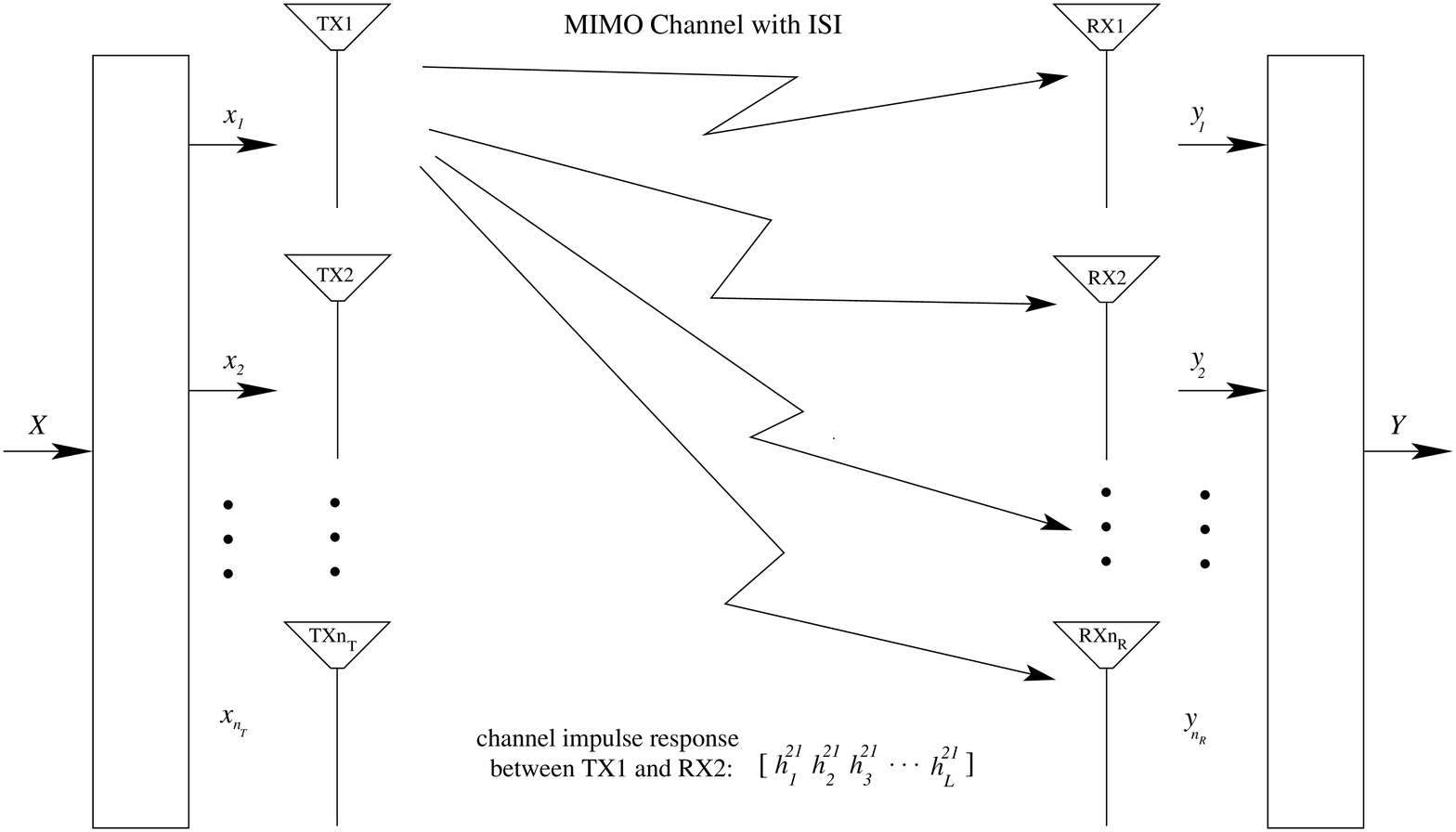}}
\caption{\label{fig:MIMOblck}Block diagram of a MIMO system with ISI.}
\end{figure}

For a MIMO wireless system with $n_T$ transmitter antenna and $n_R$ receiver antenna, the received signal at time index $n$, $Y_n=\left[y_{1,n},\cdots,y_{n_R,n}\right]^T$, will be as follows:
\begin{eqnarray}
  Y_n=HX_n+N_n,
\end{eqnarray}
where $H$ is the channel matrix, $X_n=\left[x_{1,n},\cdots,x_{n_T,n}\right]^T$ is the transmitted signal at time $n$ and $N_n$ is the noise signal. The channel matrix entries $h_{i,j}$'s reflect the channel effect on the signal transmitted  from antenna $j$ in the transmitter and received at antenna $i$ in the receiver.
In a more realistic channel modeling, one may consider the Intersymbol-Interference (ISI) \cite[Chapter 2]{Larsson-Stoica-03}. In this case, the channel impulse response between the transmitter antenna $j$ and the receiver antenna $i$ is a  vector $h_{ij}=\left[\begin{array}{ccccc}h^{(ij)}_1& h^{(ij)}_2&
\cdots&h^{(ij)}_{L-1}&h^{(ij)}_{L}\end{array}\right]^T$ where $L$ is the length of the impulse response of the channel (number of the taps). Consequently, the channel matrix  for a signal frame of $K$ will be as follows:
\begin{eqnarray}\label{eq:channel}
  H&=&\left[\
    \begin{array}{ccccccccc}
      A_1&A_2&\cdots&A_{L}&{\bf 0}&{\bf 0}&&\cdots&{\bf 0}\\
      {\bf 0}&A_1&A_2&\cdots&A_{L}&{\bf 0}&&\cdots&{\bf 0}\\
      {\bf 0}&{\bf 0}&A_1&A_2&\cdots&A_{L}&{\bf 0}&\cdots&\vdots\\
      \vdots&\vdots&\ddots&\ddots&\cdots&&\ddots&\cdots&{\bf 0}\\
      {\bf 0}&{\bf 0}&\cdots&\cdots&{\bf 0}&A_1&A_2&\cdots&A_{L}
    \end{array}
    \right],
\end{eqnarray}
where there are $K-1$ zero-matrices in each row and $A_l=(h^{\left(ij\right)}_l)_{{\scriptstyle
i=1,\cdots,n_R}\atop{\scriptstyle j=1,\cdots,n_T}}$(see Fig.~\ref{fig:MIMOblck} for the block diagram). To calculate the capacity of such a channel, one needs to know the eigenvalue distribution of the $H^*H$
\cite{Tulino-Verdu}.

Free probability and random matrix theory have proven to be
reliable tools in tackling similar problems \cite{Tulino-Verdu}. Tse and Zeitouni \cite{Tse-Zeitouni-00}
applied random matrix theory to study linear multiuser receivers, Moustakas  {\em et. al.}
\cite{Moustakas-Simon-Sengupta-03} applied it to calculate the capacity of a MIMO channel. M\"uller
\cite{Muller-02} employed it in calculating the eigenvalue distribution of a particular fading channel and
later Debbah and  M\"uller \cite{Debbah-Muller-05} applied it in MIMO channel modeling.

While the most basic random matrices are Gaussian and Wishart random matrices, and it is now quite widely
known (see, for example, \cite{Tulino-Verdu}) how to use tools from free probability (like the $R$- or
$S$-transform \cite{VDN-92}) to  deal with such random matrices, there are many interesting situations, as
the matrix $H$ above,
which cannot be modeled with these simple kind of random matrices. Very often these can, at least in first
approximation, be assumed to be block random matrices. These are  matrices built out of blocks, where each
block is a Gaussian random matrix; however, the variance of the entries might be different from block to
block and, in particular, some of the blocks might repeat at different places in the matrix.
Our main example for such block matrices is the channel matrix $H$ as in \eqref{eq:channel}, but there are
also other interesting matrices of this kind. In particular, we will also consider
block Toeplitz matrices, like the $3\times 3$ block matrix
$$X=\begin{bmatrix}
A&B&C\\
B&A&B\\
C&B&A
\end{bmatrix},$$
where $A,B,C$ are independent selfadjoint (which we use as a synonym
of Hermitian)  $N\times N$ Gaussian random matrices. Such matrices
were considered, e.g., in \cite{Tamer}, but their limiting
eigenvalue distribution for $N\to\infty$ remained open.

While M\"uller\cite {Muller-02a} has addressed a similar problem using free probability, in this paper we will show how to use a more general version of free
probability theory, so-called operator-valued free probability
theory, to calculate the asymptotic eigenvalue distribution of such
block matrices. It was the fundamental insight of Shlyakhtenko
\cite{Shlyakhtenko-96,Shlyakhtenko-97} that operator-valued free
probability can be used for dealing with special block matrices. Our
main theorems could, by building on Shlyakhtenko's work, be derived
quite directly by combining various results from the literature on
operator-valued free probability theory. However, since most readers
will not be familiar with operator-valued free probability theory,
we prefer to give also a more direct proof which does not assume any
prior knowledge on operator-valued free probability. Actually, the
"operator-valued" structure of the result will arise quite
canonically in our proof, and thus our presentation will hopefully
convince the reader not only of the power but also of the conceptual
beauty of operator-valued free probability theory and serve as an
appetizer to read more about this field.

In Section~\ref{sec:pre} some basic notations and definitions that are used 
through
the paper are introduced \cite{Nica-Speicher}. The theorems are stated in
Section~\ref{sec:statements} while they are proved in
Section~\ref{sec:proof}. In  Section~\ref{sec:results} it is shown
as a warm up exercise how the Marchenko-Pastur law can be derived
using the proposed method. Then, two general scenarios are studied
in this section. In the first scenario the eigenvalue distribution
of  block Toeplitz matrices is computed and numerical results are
shown for $3\times 3$, $4\times 4$ and $5\times 5$ block Toeplitz
matrices. The second example computes the eigenvalue distribution
for a MIMO system with Intersymbol-Interference (ISI) in two different cases.
Section~\ref{sec:results} concludes with a discussion on the
convergence and numerical behaviour of the proposed method.
Conclusion remarks are drawn in Section~\ref{sec:conclusion}.

\section{Preliminaries}\label{sec:pre}
\subsection{Notations}
The following notations are adopted in the paper:
\begin{tabbing}
test, \= jhk \= lkjd \kill
$\otimes$ \>\>$\mbox{Tensor
(Kronecker)
product of matrices}$\\
$\oplus$\>\>$\mbox{direct sum}$\\
$\Im\left(X\right)$\>\>$\mbox{Imaginary part of 
} X$\\
$I_d$\>\>$d\times d \mbox{ Identity matrix}$\\
$\overline{X}$\>\>$\mbox{complex conjugate of } X$\\
$\delta_{ij}$\>\>$\mbox{Dirac delta function}$\\
$X^*$\>\>$\mbox{Hermitian 
conjugate of matrix }X$
\end{tabbing}
\subsection{Gaussian family and Wick formula}
The entries of our random matrices will consist of Gaussian random variables; since entries might
repeat at various places in the matrix, not all entries are independent and we
use the notion of a Gaussian family to describe this general
situation. A Gaussian family is a family of random variables
$x_1,\dots,x_n$ which are linear combinations of independent
Gaussian random variables. Clearly, the whole information about the
joint distribution of such a family is contained in the covariance
matrix $C=\left(c_{ij}\right)_{i,j=1}^n$,
$E\left[x_ix_j\right]=c_{ij}$. There is a very precise combinatorial
formula, usually called Wick formula, which allows to express the
higher joint moments in a Gaussian family in terms of second moments
(i.e., in terms of the covariance matrix). Namely, for all $k\in
\mathbb{N}$ and $1\le i\left(1\right),\cdots,i\left(k\right)\le n$,
we have:
\begin{eqnarray}\label{E:Wick}
  E\left[x_{i\left(1\right)}\cdots x_{i\left(k\right)}\right]=\sum_{\pi\in\cP_2\left(k\right)}\prod_{\left(r,s\right)\in\pi}E\left[x_{i\left(r\right)}x_{i\left(s\right)}\right],
\end{eqnarray}
where $\cP_2\left(k\right)$ is the set of all pairings of the set $\left\{1,\cdots,k\right\}$.
Since there is no pairing of an odd number of elements, this formula also contains the statement that all
odd moments of a Gaussian family vanish.
\subsection{Cauchy transform and Stieltjes inversion formula}
Our main analytical object for determining the eigenvalue distribution of a random matrix is
the Cauchy transform $G_\mu$
of a probability measure $\mu$ on $\mathbb{R}$. It is defined as follows:
\begin{eqnarray}\label{def:Cauchy}
  G_\mu\left(z\right)=\int_\mathbb{R} \frac 1 {z-t} d\mu\left(z\right), \quad z\in \mathbb{C}^+,
\end{eqnarray}
where $\mathbb{C}^+:=\left\{s+it\left|s,t\in\mathbb{R}, t>0 \right.\right\}$. $G_\mu$ is analytic and takes values in $\mathbb{C}^-:=\left\{s+it\left|s,t\in\mathbb{R}, t<0 \right.\right\}$. For a compactly supported $\mu$, $G_\mu$ can be
expanded as a power series:
\begin{eqnarray}\label{E:CauchyEx}
  G_\mu\left(z\right)=\sum_{n=0}^\infty \frac{\alpha_n}{z^{n+1}}, \quad \left|z\right|>r,
\end{eqnarray}
where $r:=\sup\left\{\vert t\vert \mid t\in\mathrm{supp}\left(\mu\right)\right\}$ and
$\alpha_n:=\int_\mathbb{R}t^nd\mu\left(t\right)$ is the $n$th moment of $\mu$ for $n\ge 0$. Using
 \eqref{def:Cauchy}, it is clear that the Cauchy  transform has the following
property:
\begin{eqnarray}
  \lim_{z\in\mathbb{C}^+,\left|z\right|\rightarrow\infty}zG_\mu\left(z\right)=1.
\end{eqnarray}
One can recover the probability measure $\mu$ from its Cauchy transform by
the following formula, known as the Stieltjes inversion formula:
$$d\mu\left(t\right)=\lim_{\epsilon\rightarrow 0}h_\epsilon\left(t\right)dt$$
where
\begin{eqnarray}
  h_\epsilon\left(t\right):=-\frac 1 \pi \Im G_\mu\left(t+i\epsilon\right),\quad \forall t\in \mathbb{R}.
\end{eqnarray}

\section{Statement of the Theorems}\label{sec:statements}
\subsection{Selfadjoint block matrix with square blocks} We will consider $d\times d$-matrices with a
given block structure; the blocks themselves will be large Gaussian $N\times N$-random matrices; our aim
is
to calculate the limit eigenvalue distribution for such block matrices if the size $N$ of the
Gaussian matrices tends to infinity. The main point is that the blocks are not all different,
but they might repeat (as themselves or their adjoint) at different places. Otherwise from that
they should be independent.

We write our matrices as

\begin{equation}
  \label{Kronecker}
  X_N=\sum_{i,j=1}^d E_{ij}\otimes A^{(i,j)},
\end{equation}
where $E_{ij}$ are the elementary $d\times d$-matrices having a 1 at
the entry $(i,j)$ and 0 otherwise. For each $i,j=1,\dots,d$ we have
a Gaussian $N\times N$ random matrix $A^{(i,j)}$, 
which we call a block. In each block $A^{(i,j)}$ all the entries are
Gaussian and independent. Furthermore, we want $X_N$ to be
selfadjoint, which means that $A^{(i,j)}=A^{(j,i)*}$ for all
$i,j=1,d$. In particular, the blocks $A^{(i,i)}$ on the diagonal
must be selfadjoint. We would like to leave it open whether the
blocks $A^{(i,j)}$ with $i\not= j$ are selfadjoint or not - this
depends on the concrete situation.

A convenient way to encode the information about which blocks agree, is by specifying the
covariance $\sigma(i,j;k,l)$ between an entry in the matrix $A^{(i,j)}$ and an entry in the
matrix $A^{(k,l)}$. In the case that the two blocks are either independent or the same (where
we also include the possibility that one is the adjoint of the other), the covariance $\sigma$
takes just on the values 0 or 1. However, we can also be more general and allow arbitrary
$\sigma$'s. The only constraint comes from the requirement that $X_N$ is selfadjoint which
means that we must have
\begin{equation}
  \label{symmetry}
  \sigma(i,j;k,l)=\overline{\sigma(k,l;i,j)}\qquad
\end{equation}
for all $i,j,k,l=1,\dots,d$.

If we denote the entries of the matrix $A^{(i,j)}$ by $a^{(i,j)}_{rp}$ with $r,p=1,\dots,N$,
then we can summarize all the above by requiring that the collection of all entries
$\{a_{rp}^{(i,j)}\mid i,j=1,\dots,d,\, r,p=1,\dots,N\}$ of the matrix $X_N$ forms a Gaussian
family which is determined by
$$a^{(i,j)}_{rp}=\overline{a_{pr}^{(j,i)}}\qquad\text{for all $i,j=1,\dots,d$, $r,p=1,\dots,N$}$$
and the prescription of mean zero and
covariance
\begin{equation}\label{eq:cov-square}
E[a_{rp}^{(i,j)} a_{qs}^{(k,l)}]=\frac 1n
\delta_{rs}\delta_{pq}\cdot \sigma(i,j;k,l),
\end{equation}
where $$n=dN.$$ Note that the above gives also
the covariance between $a$'s and $\bar a$'s as
$$E[a_{rp}^{(i,j)} \overline{a_{sq}^{(l,k)}}]=\frac 1n \delta_{rs}\delta_{pq}\cdot
\sigma(i,j;k,l)$$
and thus the joint distribution of the matrix entries of $A$ is uniquely determined.

Furthermore, with this description we have the possibility of including both cases
where non diagonal blocks are selfadjoint or not (or mixtures of this).
Consider a situation where $\sigma$ takes on only the values 0 or 1.
Then a non-diagonal $A^{(i,j)}$ is selfadjoint if and only if $\sigma(i,j;i,j)=1$.
As an example, contrast the situation

$$\begin{bmatrix}
0&A_1&0\\
A_1&0&A_1\\
0&A_1&0
\end{bmatrix}$$
where $A_1$ is a selfadjoint Gaussian $N\times N$-matrix
with the situation
$$\left[\begin{matrix}
0&B_1&0\\
B_1^*&0&B_1\\
0&B_1^*&0
\end{matrix}\right],$$
where $B_1$ is a non-selfadjoint Gaussian $N\times N$-matrix.

In the first case $\sigma(i,j;k,l)=1$ for all 16 possible combinations of
$(i,j)$ and $(k,l)$ from the set $\{(1,2),(2,1),(2,3),(3,1)\}$, whereas in the second
case
$\sigma(i,j;k,l)$
is only non-vanishing for
$$\sigma(1,2;2,1),\sigma(1,2;3,2),\sigma(2,3;2,1),\sigma(2,3;3,2)$$
and the "adjoint values"
$$\sigma(2,1;1,2),\sigma(3,2;1,2),\sigma(2,1;2,3),\sigma(3,2;2,3).$$

\begin{notation}
Fix a natural number $d$ and a covariance function
$$\sigma=\bigl(\sigma(i,j;k,l)\bigr)_{i,j,k,l=1}^d$$
such that \eqref{symmetry} holds.

1)
$M_d(\CC)$ are just the $d\times
d$-matrices with complex entries,
$$M_d(\CC):=\{(d_{ij})_{i,j=1}^d \mid i,j=1,\dots,d\}.$$

2)
We define a \emph{covariance mapping}
$$\eta:M_d(\CC)\to M_d(\CC)$$ as follows:
\newline For $D=(d_{ij})_{i,j=1}^d\in M_d(\CC)$ we have $\eta(D)=(\eta(D)_{ij})_{i,j=1}^d$ with
$$\left[\eta(D)\right]_{ij}:=\frac 1d\sum_{k,l=1}^d \sigma(i,k;l,j) d_{kl}.$$

3) Furthermore, $\tr_d$ denotes the normalized trace on $M_d(\CC)$, i.e.,
$$\tr_d\bigl(D\bigr):=\frac 1d\sum_{i=1}^d [D]_{ii}.$$
\end{notation}

\begin{theorem}\label{thm:square}
With the above notation, for each $N\in\NN$, consider block matrices
\eqref{Kronecker}, where, for each $i,j=1,\dots,d$, the blocks
$A^{(i,j)}=\bigl(a^{(i,j)}_{rp}\bigr)_{r,p=1}^N$ are Gaussian
$N\times N$ random matrices such that the collection of all entries
$\{a_{rp}^{(i,j)}\mid i,j=1,\dots,d,\, r,p=1,\dots,N\}$ of the
matrix $X_N$ forms a Gaussian family which is determined by
$$a^{(i,j)}_{rp}=\overline{a_{pr}^{(j,i)}}\qquad\text{for all $i,j=1,\dots,d$, $r,p=1,\dots,N$}$$
and the prescription of mean zero and covariance
\eqref{eq:cov-square}.

 Then, for $N\to\infty$, the $n\times n$
matrix $X_N$ has a limiting eigenvalue distribution whose Cauchy
transform $G(z)$ is determined by
$$G(z)=\tr_d(\cG(z)),$$
where $\cG(z)$ is an $M_d(\CC)$-valued analytic function on the
upper complex half plane, which is uniquely determined by the facts
that
\begin{equation}\label{eq:cG lim}
\lim_{|z|\to\infty, \Im(z)>0}z\cG(z)=I_d, \end{equation}
 and that it satisfies
for all $z$ in the upper complex half plane the matrix equation
\begin{equation}\label{eq:square}
 z \cG(z)= I_d + \eta(\cG(z))\cdot \cG(z).
\end{equation}
\end{theorem}

The proof of this theorem is just a few lines if one cites the
relevant literature from operator-valued free probability theory. We
will indicate this below. However, since most readers will not be
familiar with operator-valued free probability theory, we prefer to
give in the next section a more direct proof from scratch. As one
will see, the "operator-valued" structure of the result will arise
quite canonically and the next section is also intended to introduce
the reader to some relevant notions and ideas of operator-valued
free probability theory.

Let us now give, for the reader familiar with operator-value free
probability, the condensed version of the proof of our Theorem
\ref{thm:square}. First, one has to observe that the joint moments
of the blocks converges to a semi-circular family, thus the wanted
limit distribution of $X_N$ is the same as the one of a $d\times
d$-matrix $S$, where the entries of $S$ are from a semi-circular
family, with covariance $\sigma$. By using the description of
operator-valued cumulants of this matrix in terms of the cumulants
of the entries of the matrix (see \cite{Nica-Shlahtenko-Speicher02b}
), it is obvious that $S$ is a $M_d(\CC)$-valued semi-circular
element, with covariance $\eta$. The equation for $\cG(z)$ follows
then from the basic $R$-transform or cumulant theory of
operator-valued free probability theory, see
\cite{Spe,Voi}. 

In the special case that blocks of our block matrix $X_N$ do not
repeat (apart from the symmetry condition $X_N=X_N^*$) one might
call the block matrix a band matrix (in such a situation one often lets $d$ also go
to infinity). The fundamental observation that
limits of Gaussian band matrices are operator-valued semicircular
elements -- and thus that operator-valued free probability can be
used for determining their eigenvalue distribution -- was made by
Shlyakhtenko in \cite{Shlyakhtenko-96}. There he proved actually the
special case of the above Theorem \ref{thm:square} for band
matrices. In this case, the covariance function $\eta$ maps diagonal
matrices to diagonal ones and one can consider the block matrix as a
semi-circular element over the diagonal matrices.

\subsection{Selfadjoint block matrix with rectangular blocks} In many applications one is also interested in situations
where the blocks themselves might not be square matrices, but more
general rectangular matrices. Of course, the sizes of the blocks
must fit together to make up a big square matrix. This means that we
replace $n=dN$ by a decomposition $n=N_1+\cdots +N_d$, and the block
$A^{(i,j)}$ will then be a $N_i\times N_j$-matrix. We are interested
in the limit that $N_i/n$ converges to some number $\alpha_i$. It
turns out that one can modify our Theorem \ref{thm:square} (and also
its proof) quite easily from the square to the general rectangular
situation. Let us first introduce the generalizations of our
relevant notations from the square case. Note that dependent
rectangular blocks can be re-cut into different nonequivalent
configurations of dependent blocks. We will assume that such
repartitioning has already been done and resulted in
 the covariance function $\sigma(i,j;k,l)$ that can only be different from zero if
the size of the block $A^{(i,j)}$ fits (at least in the limit $n\to\infty$) with the size of
the block $A^{(k,l)}$.

\begin{notation}
Fix a natural number $d$ and a $d$-tuple
$\alpha=(\alpha_1,\dots,\alpha_d)$ with $0<\alpha_i<1$ for all
$i=1,\dots,d$ and $\alpha_1+\cdots+\alpha_d=1$. Furthermore, let a
covariance function
$\sigma=\bigl(\sigma(i,j;k,l)\bigr)_{i,j,k,l=1}^d$ be given with the
property that
\eqref{symmetry} holds and in addition
$\sigma(i,j;k,l)=0$ unless $\alpha_i=\alpha_l$ and
$\alpha_j=\alpha_k$. Then we use the following notations.

1) $M_\alpha(\CC)$ are those matrices from $M_d(\CC)$ which correspond to square blocks,
$$M_\alpha(\CC):=\{(d_{ij})\in M_d(\CC)\mid
\text{$d_{ij}=0$ unless $\alpha_i=\alpha_j$}\}.$$

2) We define the \emph{weighted covariance mapping}
$$\eta_\alpha:M_\alpha(\CC)\to M_\alpha(\CC)$$ as follows:
\newline For $D=(d_{ij})_{i,j=1}^d\in M_\alpha(\CC)$ we have 
$\eta_\alpha(D)=(\eta_\alpha(D)_{ij})_{i,j=1}^d$ with
$$\left[\eta_\alpha(D)\right]_{ij}:=\sum_{k,l=1}^d \sigma(i,k;l,j)\alpha_{k} d_{kl}.$$

3) Furthermore, the weighted trace
$$\tr_\alpha: M_\alpha(\CC)\to M_\alpha(\CC)$$
is given by
$$\tr_\alpha\bigl((d_{ij})_{i,j=1}^d\bigr):=\sum_{i=1}^d \alpha_i d_{ii}.$$

\end{notation}

\begin{theorem}\label{thm:rectangular}
With the above notation, for
$\left\{N_1,\dots,N_d\right\}\subset\NN$ consider block matrices
$$X_{N_1,\dots,N_d}=\sum_{i,j=1}^d E_{ij}\otimes A^{(i,j)}.$$
Here $E_{ij}$ are the elementary $d\times d$-matrices having a 1 at
the entry $(i,j)$ and 0 otherwise and, for each $i,j=1,\dots,d$, the
$A^{(i,j)}$ are Gaussian $N_i\times N_j$ random matrices,
$A^{(i,j)}=\bigl(a^{(i,j)}_{rp}\bigr)_{r=1,\dots,N_i\atop
p=1,\dots,N_j}$. The latter are such that the collection of all
entries $\{a_{rp}^{(i,j)}\mid i,j=1,\dots,d,\, r=1,\dots,N_i,
p=1,\dots,N_j\}$ of the matrix $X_{N_1,\dots,N_d}$ forms a Gaussian
family which is determined by
$$a^{(i,j)}_{rp}=\overline{a_{pr}^{(j,i)}}\qquad\text{for all $i,j=1,\dots,d$, $r=1,\dots,N_i$,
$p=1,\dots,N_j$}$$ and the prescription of mean zero and covariance
$$E[a_{rp}^{(i,j)} a_{qs}^{(k,l)}]=\frac 1n \delta_{rs}\delta_{pq}\cdot
\sigma(i,j;k,l),$$
where we put
$$n:=N_1+\cdots N_d.$$
Then, for $n\to\infty$ such that
$$\lim_{n\to\infty} \frac {N_i}n=\alpha_i\qquad\text{for all $i=1,\dots,d$},$$
the matrix $X_{N_1,\dots,N_d}$ has a limiting eigenvalue
distribution whose Cauchy transform $G(z)$ is determined by
$$G(z)=\tr_\alpha(\cG(z)),$$
where $\cG(z)$ is an $M_\alpha(\CC)$-valued analytic function on the
upper complex half plane, which is uniquely determined by the facts
that \eqref{eq:cG lim} holds and that it satisfies for all $z$ in
the upper complex half plane the matrix equation
\begin{equation} \label{eq:rectangular}
z \cG(z)= I_d + \eta_\alpha(\cG(z))\cdot \cG(z).
\end{equation}
\end{theorem}

In principle, the statement for the rectangular situation can be
reduced to the case of square blocks by cutting the rectangular
blocks into smaller square blocks (at least asymptotically); thus
Theorem \ref{thm:rectangular} can be deduced from Theorem
\ref{thm:square}. However, we find it more instructive to adapt our
proof of Theorem \ref{thm:square} to the rectangular situation -
this makes the structure of the weighted covariance mapping and the
appearance of the weighted trace more transparent. We will present
this proof of Theorem \ref{thm:rectangular} also in the next
section. Let us also mention that the considerations of Shlyakhtenko
for Gaussian band matrices were extended to rectangular matrices by
Benaych-Georges \cite{Benaych-Georges-05}; thus some of his results
are closely related to the special case of band matrices in our
Theorem \ref{thm:rectangular}.

\subsection{Wishart type block matrices}
Another type of block matrices appearing in applications is of a
Wishart type. Here the block matrix $H$ itself is rectangular and
one is interested in the eigenvalue distribution of $HH^*$. Let us
write $H=(A^{(i,j)})_{i=1,\dots,r\atop j=1,\dots,s}$, where each
block $A^{(i,j)}$ is a Gaussian $M_i\times N_j$ random matrix. Put
$M:=M_1+\cdots+M_r$ and $N:=N_1+\cdots N_s$, so $H$ is an $M\times
N$-matrix.  The calculation of the eigenvalue distribution of $HH^*$
can be reduced to the situation treated in the previous section by
the following trick. Consider
\begin{equation}\label{eq:X}X=\begin{bmatrix}
0& H\\
H^*&0
\end{bmatrix}.\end{equation}
With $n=M+N$, this is a selfadjoint $n\times n$-matrix and can be
viewed as a $d\times d$-block matrix of the form considered in the
previous section, where $n=M+N$ is decomposed into $d:=r+s$ parts
according to
$$M+N=M_1+\cdots+M_r+N_1+\cdots N_s.$$
Thus we can use Theorem \ref{thm:rectangular} to get the eigenvalue
distribution of $X$ if the $M_i$ and $N_j$ go to infinity with a
limit for their ratio.

The only remaining question is how to relate the eigenvalues of $X$ with those of $HH^*$. This
is actually quite simple, we only have to note that all the odd moments of $X$ are zero and
$$X^2=\begin{bmatrix}
HH^*&0\\
0& H^*H
\end{bmatrix}$$
Thus the eigenvalues of $X^2$ are the eigenvalues of $HH^*$ together
with the eigenvalues of $H^*H$. Furthermore, note that $HH^*$ is an
$M\times M$ and $H^*H$ is an $N\times N$ matrix. Assuming that $M<N$
(otherwise exchange the role of $H$ and $H^*$) we have then that the
eigenvalues of $H^*H$ are the eigenvalues of $HH^*$ plus $N-M$
additional zeros. In terms of the unnormalized trace $\Tr$ this just
means
$$\Tr(X^{2k})=\Tr((HH^*)^k)+\Tr((H^*H)^k)=2\Tr ((HH^*)^k)$$
Rewritten in normalized traces this gives 
$$\tr_{n}(X^{2k})=\frac {2M}{n}\tr_M\left((HH^*)^k\right).$$
Putting all these together, it results in the following relation between the Cauchy transform
$G_{X^2}$ of $X^2$ and the Cauchy transform $G_{HH^*}$ of $HH^*$:
\begin{equation}\label{eq:HHst-X2}
G_{HH^*}(z)=\frac{M+N}{2M} G_{X^2}(z)-\frac {N-M}{2M} \frac 1z.
\end{equation}
Finally, we should also rewrite our equation for the Cauchy transform $G_X$ of $X$
in terms of $G_{X^2}$.
Since $X$ is even, both are related by
$$z\cdot G_{X^2}(z^2)=G_X(z).$$
By noting that the operator-valued version $\cG(z)$ depends only on
$z^2$, we can introduce a quantity $\cH$ by
$$z\cdot\cH(z^2)=\cG(z).$$
Then we have
\begin{equation}\label{eq:eins}
\lim_{n\to\infty} G_{X^2}(z)=\tr_\alpha[\cH(z)],
\end{equation}
and the equation (\ref{eq:square}) for $\cG$ becomes
\begin{equation}\label{eq:zwei}
z \cH(z)= I_d+z\eta\bigl(\cH(z)\bigr)\cdot \cH(z).
\end{equation}
The equations (\ref{eq:eins}), (\ref{eq:zwei}), together with
(\ref{eq:HHst-X2}) determine the asymptotic eigenvalue distribution
of $HH^*$ by
\begin{equation}
  \label{eq:X2H}
  \lim_{M+N\to\infty}G_{HH^*}(z)=\theta \tr_{\alpha}\cH(z)-\frac{\theta_0}z,
\end{equation}
where
$$\theta=\lim_{N\to\infty}\frac{M+N}{2M}=\frac{1}{2\sum_{j=1}^r\alpha_j},\;
\theta_0=\lim_{N\to\infty}\frac{N-M}{2M}=\frac{\sum_{j=r+1}^d
\alpha_j-\sum_{j=1}^r\alpha_j}{2\sum_{j=1}^r\alpha_j}.$$ Note that
the subtraction of a pole at 0 in (\ref{eq:HHst-X2}) reflects just
the occurrence of spurious zeros in $X^2$ which are coming from
$H^*H$.

It would seem more adequate to have access to the information about
the eigenvalues of $HH^*$ without having to invoke the eigenvalue
distribution of $H^*H$. To simplify notations we will restrict in
the following to the situation where all blocks $A^{(i,j)}$ are
square matrices i.e.,
 $N_1=\dots=N_r=M_1=\dots=M_s$, and thus $N=rN_1$
and $M=sN_1$. (Note that $H$ itself might still be rectangular,
i.e.,
 we are allowing that $r\not=s$.) One
can of course write our $(r+s)\times (r+s)$ matrix $\cH$ as a
$2\times 2$-block matrix
$$\cH(z)=\begin{bmatrix}
\cG_1(z)& \cG_3(z)\\
\cG_3^*(z)& \cG_2(z)
\end{bmatrix}$$
where $\cG_1(z)$, $\cG_2(z)$, and $\cG_3(z)$ are $r\times r$-,
$s\times s$-, and $r\times s$-matrices, respectively. From our
considerations in the next section is fairly easy to see that
\begin{equation}\label{eq:vier}
\lim_{n\to\infty}G_{HH^*}(z)=\tr_r\bigl(\cG_1(z)\bigr).
\end{equation}
One should note that the quadratic matrix equation (\ref{eq:zwei})
for $\cH$ will in general couple all the various $\cG_i$, so that
even in this formulation the calculation of the distribution of
$HH^*$ will still involve $H^*H$. However, there are special
situations where one can eliminate $H^*H$ quite easily from our
equations. We will examine one such situation, which appears in most
applications to wireless communications, in more detail in the rest
of this section.

Let us consider the special version of our Wishart type block
matrices where the blocks of $H$ themselves are non-selfadjoint
Gaussian random matrices (which are still, as before, square
matrices). In this case the covariance $\sigma$ for the blocks of
\eqref{eq:X} couples effectively only blocks from $H$ with blocks
from $H^*$ and thus $\eta:M_{r+s}(\CC)\to M_{r+d}(\CC)$ maps
$M_r(\CC)\oplus M_s(\CC)$ to itself by
$$\eta:\begin{bmatrix}
D_1&0\\
0&D_2
\end{bmatrix}\mapsto
\begin{bmatrix}
\eta_2(D_2)&0\\
0&\eta_1(D_1)
\end{bmatrix},
$$
where
$$\eta_1:M_r(\CC)\to M_s(\CC) \qquad\text{and}\qquad
\eta_2:M_s(\CC)\to M_r(\CC).$$ One sees now easily that $\cH$ must
take on values in $M_r(\CC)\oplus M_s(\CC)$, thus
$$\cH(z)=\begin{bmatrix}
\cG_1(z)&0\\
0&\cG_2(z)
\end{bmatrix},
$$
where $\cG_1$ and $\cG_2$ are $M_r(\CC)$-valued and
$M_s(\CC)$-valued, respectively, analytic functions in the upper
complex half plane. The equation (\ref{eq:zwei}) for $\cH$ splits
now into the two equations
$$z\cG_1(z)=
I_r+z\eta_2\bigl(\cG_2(z)\bigr)\cdot \cG_1(z)$$ and
$$z\cG_2(z)=
I_s+z\eta_1\bigl(\cG_1(z)\bigr)\cdot \cG_2(z).$$ One can eliminate
$\cG_2$ from those equations by solving the second equation for
$\cG_2$ and inserting this into the first equation, yielding
$$z\cG_1(z)=
I_r+ \eta_2\Bigl(\bigl( I_s-\eta_1\bigl(\cG_1(z)\bigr)\bigr)^{-1}
\Bigr)\cdot \cG_1(z)$$ Together with (\ref{eq:vier}) this gives
directly the eigenvalue distribution of $HH^*$.

\section{Proof of the Main Theorems}\label{sec:proof}
\subsection{Proof of Theorem \ref{thm:square}}
We will prove Theorem 2 by calculating the moments of the averaged
eigenvalue distribution of the $n\times n$ random matrices $X_N$ in
the limit $N\to\infty$. Note that the $m$-th moment of the averaged
eigenvalue distribution is just given by the expectation of the
normalized trace $\tr_n$ of the $m$-th power of $X_N$. For this we
can calculate
\begin{align*}
E[\tr_n(X_N^m)]&=\frac 1{d}\sum_{i(1),\dots,i(m)=1}^d E\bigl[
\tr_N(A^{(i(1),i(2))}\cdots A^{(i(m),i(1))})\bigr]\\
&= \frac 1{dN} \sum_{i(1),\dots,i(m)=1}^d \sum_{r(1),\dots,r(m)=1}^N
E\bigl[ a_{r(1)r(2)}^{(i(1),i(2))}\cdots
a_{r(m)r(1)}^{(i(m),i(1))}\bigr]
\end{align*}
Now we need to invoke the Wick formula for calculating the
expectation of a product of entries from our matrix. The Wick
formula allows to express all higher moments in random variables
from a Gaussian family in terms of the covariances by a nice
combinatorial formula, by summing over all pairings of the appearing
factors. In our case this gives
\begin{align*}
&E\bigl[ a_{r(1)r(2)}^{(i(1),i(2))}\cdots
a_{r(m)r(1)}^{(i(m),i(1))}\bigr] =\sum_{\pi\in\cP_2(m)}
\prod_{(p,q)\in\pi} E[a_{r(p)r(p+1)}^{(i(p),i(p+1))}
a_{r(q)r(q+1)}^{(i(q),i(q+1))}].
\end{align*}
Here $\cP_2(m)$ denotes the pairings of a set of $m$ elements. A
pairing is just a decomposition into subsets (called blocks of the
pairing) with two elements. The product in the above Wick formula
runs over all blocks of the pairing $\pi$. If we now use the formula
(\ref{eq:cov-square}) for the covariances of the entries of our
random matrix $X_N$, then we can continue with
\begin{align*}
&E\bigl[ a_{r(1)r(2)}^{(i(1),i(2))}\cdots
a_{r(m)r(1)}^{(i(m),i(1))}\bigr]
\\
&\quad=\sum_{\pi\in\cP_2(m)} \prod_{(p,q)\in\pi} \frac 1n\cdot
\sigma \bigl(i(p),i(p+1);i(q),i(q+1)\bigr)\cdot
\delta_{r(p)r(q+1)}\cdot \delta_{r(p+1)r(q)}.
\end{align*}
In order to understand better the contribution of a pairing $\pi$
according to this formula, it will be useful to interpret this
formula in the language of permutations. By $S_m$ we will denote the
group of permutations on $m$ elements. Let us denote by $\gamma\in
S_m$ the cyclic permutation $\gamma=(1,2,\dots,m)$. If we identify a
partition $\pi\in \cP_2(m)$ with a permutation in $S_m$ -- by
declaring the blocks to cycles, i.e., $\pi(p)=q$ and $\pi(q)=p$ for
$(p,q)\in\pi$ -- then one recognizes easily that
$$\sum_{r(1),\dots,r(m)=1}^N  \prod_{(p,q)\in\pi} \delta_{r(p)r(q+1)}\cdot \delta_{r(p+1)r(q)}
=\sum_{r(1),\dots,r(m)=1}^N \prod_{p=1}^m \delta_{r(p)
r(\gamma\pi(p))}=N^{\#\gamma\pi},$$ where $\#\gamma\pi$ denotes the
number of cycles of the permutation $\gamma\pi$.

With this we can continue our above calculation of the expected
trace of $X_N^m$ as follows:
$$E[\tr_n(X_N^m)]=\sum_{\pi\in\cP_2(m)} n^{\#\gamma\pi-m/2-1}
\frac 1{d^{\#\gamma\pi}}\sum_{i(1),\dots,i(m)=1}^d
\prod_{(p,q)\in\pi}
 \sigma
\bigl(i(p),i(p+1);i(q),i(q+1)\bigr)$$

Note that for $d=1$ (and $\sigma(1,1;1,1)=1$) we recover the
well-known genus expansion
$$E[\tr_n(A^m)]=\sum_{\pi\in\cP_2(m)} n^{\#\gamma\pi-m/2-1}$$
of a selfadjoint Gaussian $n\times n$ random matrix.

It is quite easy to see that
$$\#\gamma\pi-m/2-1\leq 0\qquad\text{for any $\pi\in\cP_2(m)$},$$
and that we have equality there exactly for the so-called
non-crossing pairings $\pi\in NC_2(m)$. The latter are those
pairings $\pi$ of $1,2,\dots,2m$ for which the blocks of $\pi$ can
be drawn in such a way below the numbers that they do not cross.
Another useful way of describing non-crossing pairings is the
following recursive procedure: A pairing of $2m$ numbers is
non-crossing if and only if one can find a block consisting of two
consecutive numbers such that after removing this block one gets a
non-crossing pairing of the $2m-2$ remaining numbers.  For
illustration, here are the 5 of the 15 pairings of 6 elements which
are non-crossing:

$$
\usebox{\NCseZ}\qquad\usebox{\NCszZ}\qquad
\usebox{\NCsdZ}
$$
and
$$
\usebox{\NCsvZ}\qquad
\usebox{\NCsfZ}
$$

So in the limit, $N\to\infty$, only the non-crossing pairings survive
in our formula and we get
$$\lim_{N\to\infty} E[\tr_n(X_N^m)]=\sum_{\pi\in NC_2(m)} {\mathcal{K}}_\pi,$$
where
$$
\KK_\pi:=\frac 1{d^{m/2+1}} \sum_{i(1),\dots,i(m)=1}^d
\prod_{(p,q)\in\pi}
 \sigma
\bigl(i(p),i(p+1);i(q),i(q+1)\bigr)$$

In the case $d=1$, all $\KK_\pi$ would be equal and we would recover
the moments of a semicircular distribution (in the form of the
Catalan numbers counting non-crossing pairings). This would
reproduce more or less the derivation of Wigner of his famous
semicircle law. For $d>1$, however, the $\KK_\pi$ are in general
different for different $\pi$, and, even worse, there does not exist
a straightforward recursive way of expressing $\KK_\pi$ in terms of
"smaller" $\KK_\sigma$. Thus we are outside the realm of the usual
techniques of free probability theory.

However, one can save most of those techniques by going over to an
"operator-valued" level. The main point of such an operator-valued
approach is to write $\KK_\pi$ as the trace of a $d\times d$-matrix
$\kk_\pi$, and then realize that $\kk_\pi$ has the usual nice
recursive structure of semicircular elements.

Namely, let us define a matrix
$$\kk_\pi=(\kk_\pi(i,j))_{i,j=1}^d$$
by
$$
\kk_\pi(i,j):= \sum_{i(1)\dots,i(m),i(m+1)=1}^d
\delta_{ii(1)}\delta_{ji(m+1)} \prod_{(p,q)\in\pi}\frac 1d\cdot
 \sigma
\bigl(i(p),i(p+1);i(q),i(q+1)\bigr).$$ Then clearly we have
$$\KK_\pi=\tr_d (\kk_\pi).$$
Furthermore, the value of $\kk_\pi$ can be determined by an iterated
application of our covariance mapping $\eta:M_d(\CC)\to M_d(\CC)$.
Namely, the value of $\kk_\pi$ is given by an iterated application
of this mapping $\eta$ according to the nesting of the blocks of
$\pi$. If one identifies a non-crossing pairing with a putting of
brackets, then the way that $\eta$ has to be iterated is quite
obvious. Let us clarify these remarks with an example.

\setlength{\unitlength}{0.4cm}

Consider the non-crossing pairing
$$\pi=\{(1,4),(2,3),(5,6)\}\in NC_2(6)\qquad\entspricht
\begin{picture}(6,2)
\thicklines \put(0,0){\line(0,1){2}} \put(0,0){\line(1,0){3}}
\put(3,0){\line(0,1){2}} \put(1,1){\line(0,1){1}}
\put(1,1){\line(1,0){1}} \put(2,1){\line(0,1){1}}
\put(4,0){\line(0,1){2}} \put(4,0){\line(1,0){1}}
\put(5,0){\line(0,1){2}}
\end{picture}
$$

The corresponding $\kk_\pi$ is
\begin{multline*}
\kk_\pi(i,j)= \frac 1{d^3}\sum_{i(2),i(3),i(4),i(5),i(6)=1}^d
\sigma(i,i(2);i(4),i(5))\cdot \sigma(i(2),i(3);i(3),i(4))\\ \cdot
\sigma(i(5),i(6);i(6),j).
\end{multline*}

$\pi$ is a non-crossing pairing of the numbers
$\mathbf{1,2,3,4,5,6}$. Let us put the indices $i=i(1),
i(2),i(3),i(4),i(5),i(6), i(7)=j$ between those numbers. Each block
of $\pi$ corresponds to one factor $\sigma$ in $\kk_\pi$ and the
arguments in this $\sigma$ are the  indices to the left and to the
right of the two numbers which are paired by this block.

\begin{picture}(6,4)
\thicklines \put(0,0){\line(0,1){2}} \put(0,0){\line(1,0){15}}
\put(15,0){\line(0,1){2}} \put(5,1){\line(0,1){1}}
\put(5,1){\line(1,0){5}} \put(10,1){\line(0,1){1}}
\put(20,0){\line(0,1){2}} \put(20,0){\line(1,0){5}}
\put(25,0){\line(0,1){2}} \put(2,2){i(2)} \put(7,2){i(3)}
\put(12,2){i(4)} \put(17,2){i(5)} \put(22,2){i(6)} \put(27,2){j}
\put(-2,2){i} \put(-0.2,2.5){\bf{1}} \put(4.8,2.5){\bf{2}}
\put(9.8,2.5){\bf{3}} \put(14.8,2.5){\bf{4}} \put(19.8,2.5){\bf{5}}
\put(24.8,2.5){\bf{6}}
\end{picture}

As mentioned before, for a non-crossing pairing one always has at
least one block consisting of neighbours; in our case such a block
is $(\mathbf{2,3})\in\pi$ (actually $(\mathbf{5,6})$ is another
one). This means that we can sum over the corresponding index $i(3)$
without interfering with other blocks, giving
\begin{align*}
\kk_\pi(i,j) &=\frac 1{d^2}\sum_{i(2),i(4),i(5),i(6)=1}^d
\sigma(i,i(2);i(4),i(5))\cdot
 \sigma(i(5),i(6);i(6),j)\\
&\qquad\qquad\qquad\qquad\qquad\qquad\qquad\qquad\qquad \cdot\frac
1d
\sum_{i(3)=1}^d \sigma(i(2),i(3);i(3),i(4))\\
&=\frac 1 {d^2}\sum_{i(2),i(4),i(5),i(6)=1}^d \sigma(i,i(2);i(4),i(5))\cdot
 \sigma(i(5),i(6);i(6),j)\cdot \eta(1)_{i(2)i(4)}
\end{align*}

Effectively we have removed the block $(\mathbf{2,3})$ and replaced
it by the matrix $\eta(I_d)$; we are left with

\begin{picture}(6,4)
\thicklines \put(0,0){\line(0,1){2}} \put(0,0){\line(1,0){15}}
\put(15,0){\line(0,1){2}} \put(20,0){\line(0,1){2}}
\put(20,0){\line(1,0){5}} \put(25,0){\line(0,1){2}} \put(2,2){i(2)}
\put(5.8,1){$\left[\eta(I_d)\right]_{i(2)i(4)}$} \put(12,2){i(4)}
\put(17,2){i(5)} \put(22,2){i(6)} \put(27,2){j} \put(-2,2){i}
\put(-0.2,2.5){\bf{1}} \put(14.8,2.5){\bf{4}} \put(19.8,2.5){\bf{5}}
\put(24.8,2.5){\bf{6}}
\end{picture}

Now the block $(\mathbf{1,4})$ of $\pi$ has become a block
consisting of neighbours and we can do the summation over $i(2)$ and
$i(4)$ without interfering with the other blocks, thus yielding
\begin{align*}
\kk_\pi(i,j)
&=\frac 1d\sum_{i(5),i(6)=1}^d  \sigma(i(5),i(6);i(6),j)\\
&\qquad\qquad\qquad\qquad \cdot\frac
1d\sum_{i(2),i(4)=1}^d\sigma(i,i(2);i(4),i(5))\cdot
\left[\eta(I_d)\right]_{i(2)i(4)}\\
&=\frac 1 d\sum_{i(5),i(6)=1}^d  \sigma(i(5),i(6);i(6),j) \cdot
\left[\eta\bigl(\eta(I_d)\bigr)\right]_{ii(5)}
\end{align*}
We have now removed the block $(\mathbf{1,4})$ and the effect of
this was that we had to apply $\eta$ to whatever was embraced by
this block (in our case, $\eta(I_d)$). We remain with

\begin{picture}(6,4)
\thicklines \put(12,0){\line(0,1){2}} \put(12,0){\line(1,0){5}}
\put(17,0){\line(0,1){2}}
\put(2,1){$\left[\eta\bigl(\eta(I_d)\bigr)\right]_{ii(5)}$}
\put(9,2){i(5)} \put(14,2){i(6)} \put(19,2){j} \put(-2,2){i}
\put(11.8,2.5){\bf{5}} \put(16.8,2.5){\bf{6}}
\end{picture}

Now we can do the summation over $i(5)$ and $i(6)$ corresponding to
the last block $(\mathbf{5,6})$ of $\pi$, which results in
\begin{align*}
\kk_\pi(i,j) &=\sum_{i(5)=1}^d
\left[\eta\bigl(\eta(I_d)\bigr)\right]_{ii(5)}\bigl(\frac
1d\sum_{i(6)=1}^d \sigma(i(5),i(6);i(6),j)
\bigr)\\
&=\sum_{i(5)=1}^d
\left[\eta\bigl(\eta(I_d)\bigr)\right]_{ii(5)}\cdot
\eta(1)_{i(5)j}\\
&=\left[\eta\bigl(\eta(I_d)\bigr)\cdot \eta(I_d)\right]_{ij}
\end{align*}
Thus we finally have
$$\kk_\pi=\eta\bigl(\eta(I_d)\bigr)\cdot \eta(I_d),$$
which corresponds to the bracket expression
$$(X(XX)X)(XX).$$
In the same way every non-crossing pairing results in an iterated
application of the mapping $\eta$. For the five non-crossing
pairings of six elements one gets the following results:
$$
\begin{matrix}
\text{\usebox{\NCseZ}\qquad\qquad\qquad}&\text{\usebox{\NCszZ}}\qquad\qquad\qquad
\\
\eta(I_d)\cdot \eta(I_d)\cdot \eta(I_d)& \eta(I_d)\cdot
\eta\bigl(\eta(I_d)\bigr)
\end{matrix}
$$
\vskip1cm
$$
\begin{matrix}
\text{\usebox{\NCsdZ}}\qquad\qquad\qquad&\text{\usebox{\NCsvZ}}\qquad\qquad\qquad\\
\eta\bigl(\eta(I_d)\bigr)\cdot\eta(I_d)&
\eta\bigl(\eta(I_d)\cdot\eta(I_d)\bigr)
\end{matrix}
$$
and \vskip1cm
$$
\begin{matrix}
\text{\usebox{\NCsfZ}}\qquad\qquad\qquad\\
\eta\Bigl(\eta\bigl(\eta(I_d)\bigr)\Bigr)
\end{matrix}
$$

Thus for $m=6$ we get
\begin{multline*}
\lim_{n\to\infty}E[\tr(A^6)]\\
=\tr_d\Bigl\{ \eta(I_d)\cdot \eta(I_d)\cdot \eta(I_d)+
\eta(I_d)\cdot \eta\bigl(\eta(I_d)\bigr)+
\eta\bigl(\eta(I_d)\bigr)\cdot\eta(I_d)+
\eta\bigl(\eta(I_d)\cdot\eta(I_d)\bigr)+ \eta\bigl(\eta
\bigl(\eta(I_d)\bigr)\bigr)\Bigr\}.
\end{multline*}

Let us summarize our calculations for general moments: We have
$$\lim_{N\to\infty} E[\tr_{n}(X_N^m)]=\tr_d\bigl\{ \sum_{\pi\in NC_2(m)} \kk_\pi\bigr\},$$
where $\kk_\pi$ are $d\times d$ matrices, determined in a recursive
way as above, given by an iterated application of the mapping
$\eta$.

Let us denote by $\EE$ the limiting operator-valued moments of
$X_N$, i.e.,
$$\EE(X^m):=\sum_{\pi\in NC_2(m)} \kk_\pi,$$
so that we have
$$\lim_{N\to\infty} E[\tr_{n}(X_N^m)]=\tr_d( \EE(X^m)).$$

An element $X$ whose operator-valued moments $\EE(X^m)$ are
calculated in such a way is called an \emph{operator-valued
semicircular element}.

The above description of the moments $\EE(X^m)$ can, similarly as
for an ordinary semicircular variable, be reformulated in a
recursive way which leads directly to our equation
(\ref{eq:square}). To get this recursive description, one has to
observe that if $\pi$ is a non-crossing pairing of $m$ elements, and
$(1,r)$ is the block of $\pi$ containing 1, then the remaining
blocks of $\pi$ must fall into two classes, those making up a
non-crossing pairing of the numbers $2,3,\dots,r-1$ and those making
up a non-crossing pairing of the numbers $r+1,r+2,\dots,m$. Let us
call the former pairing $\pi_1$ and the latter $\pi_2$, so that we
can write $\pi=(1,r)\cup \pi_1\cup \pi_2$. Then the above
description of $k_\pi$ shows that
$$\kk_\pi=\eta(\kk_{\pi_1})\cdot \kk_{\pi_2}.$$
This results for the operator valued moments in the recurrence
relation
$$\EE[X^m]=\sum_{k=0}^{m-2} \eta\bigl(\EE[X^k]\bigr)\cdot \EE[X^{m-k-2}].$$
If we go over to the corresponding generating power series,
$$\cM(z)=\sum_{m=0}^\infty \EE[X^m]z^m,$$
then this yields the relation
$$\cM(z)=I_d+z^2 \eta\bigl(\cM(z)\bigr)\cdot \cM(z).$$
Note that $M(z):=\tr_d(\cM(z))$ is the generating power series of
the moments $E[\tr_N(X_N^m)]$ for $N\to\infty$, thus it is
preferable to go over from $\cM(z)$ to the corresponding
operator-valued Cauchy transform
$$\cG(z):=\frac 1z\cM(1/z).$$
For this the above equation takes on the form
$$z \cG(z)= I_d + \eta(\cG(z))\cdot \cG(z),$$
which is exactly our equation (\ref{eq:square}). Furthermore, we
have for the Cauchy transform $G$ of the limit eigenvalue
distribution of our block matrix $X_N$ that
$$G(z)=\frac {M(1/z)}z=\tr_d(\frac {\cM(1/z)}z)=\tr_d (\cG(z)).$$

Let us finally make a few remarks about the support of the limiting
eigenvalue distribution and the validity of equation
(\ref{eq:square}) for all z in the upper half plane.

First note that the number of non-crossing pairings of $2k$ elements
is given by the Catalan numbers which behave asymptotically for
large $k$ like $4^k$. (Of course, for $m$ odd there are no pairings of $m$ elements
at all.) This implies that we can estimate the
(operator) norm of the matrix $\EE(X^m)$ by
$$\Vert \EE(X^m)\Vert \leq \Vert \eta\Vert ^m\cdot \# NC_2(m)\leq c\cdot\Vert\eta\Vert^m \cdot 2^m,$$
where $c$ is a constant independent of $m$. Applying $\tr_d$, this
yields that the support of the limiting eigenvalue distribution of
$X_N$ is contained in the interval $[-2\Vert
\eta\Vert,+2\Vert\eta\Vert]$. Clearly, since all odd moments are
zero, the measure is symmetric. Furthermore, the above estimate on
the operator-valued moments $\EE(X^m)$ shows that
$$\cG(z)=\sum_{k=0}^\infty \frac{\EE(X^{2k})}{z^{2k+1}}$$
is a power series expansion in $1/z$ of $\cG(z)$, which converges in
a neighborhood of $\infty$. Since the mapping
$$D\mapsto \frac 1zI_d +\frac 1z \eta(D)\cdot D$$
is a contraction for $\vert z\vert$ sufficiently large, $\cG(z)$ is,
for large $z$, uniquely determined as the solution of the equation
(\ref{eq:square}).

In order to realize that $\cG(z)$ can actually be extended to an
analytic matrix-valued function on the whole upper complex half
plane one has to note that the equation
\begin{multline*}
\lim_{n\to\infty} E[\tr_n(X_N^m)]\\=\sum_{\pi\in NC_2(m)} \frac
1{d^{m/2+1}}\sum_{i(1),\dots,i(m)=1}^d \prod_{(p,q)\in\pi}
 \sigma
\bigl(i(p),i(p+1);i(q),i(q+1)\bigr)
\end{multline*}
can also be read in the way that the limiting moments of $X_N$ are
given by the moments of a $d\times d$-matrix
$$X=(c_{ij})_{i,j=1}^d,$$
where the entries of $X$ form a circular/semicircular family living
in some non-commutative probability space $(\cA,\ff)$, with
covariance $\ff(c_{ij}c_{kl})=\sigma(i,j;k,l)$:
$$\lim_{n\to\infty} E[\tr_n(X_N^m)]\\=\ff\otimes \tr_d (X^m).$$
What we showed above was then essentially that $X$ is an
operator-valued semicircular variable over $M_d(\CC)$, where the
expectation
$$\EE:M_d(\cA)=M_d(\CC)\otimes\cA\to M_d(\CC)$$
is given by applying $\ff$ entrywise, i.e., $\EE=1\otimes \ff$. In
this language,
$$\cG(z)=\EE\bigl(\frac 1{z-X}\bigr),$$
which shows that it is actually an analytic function on the whole
upper complex half plane. The validity of (\ref{eq:square}) for all
$z$ in the upper half plane follows then by analytic continuation.

\subsection{Proof of Theorem \ref{thm:rectangular}}
The proof of Theorem \ref{thm:rectangular} is very similar to the
one of Theorem \ref{thm:square}. We will concentrate here mainly on
the modifications compared to the latter one.

Again, one calculates the averaged traces of powers by invoking the
Wick formula. This yields
\begin{align*}
&E[\tr_n(X_{N_1,\dots,N_d}^m)]=\frac 1{n} \sum_{i(1),\dots,i(m)=1}^d
\,\sum_{r(1)=1}^{N_{i(1)}}\dots\sum_{r(m)=1}^{N_{i(m)}}
\sum_{\pi\in\cP_2(m)} \prod_{(p,q)\in\pi} \frac 1n\\
&\qquad\qquad\qquad \cdot \sigma
\bigl(i(p),i(p+1);i(q),i(q+1)\bigr)\cdot \delta_{r(p)r(q+1)}\cdot
\delta_{r(p+1)r(q)}\\
&= \sum_{\pi\in\cP_2(m)}
n^{\#\gamma\pi-m/2-1}\sum_{i(1),\dots,i(m)=1}^d
\prod_{c\in\gamma\pi} \frac{N_{i(c)}}{n} \prod_{(p,q)\in\pi}
 \sigma
\bigl(i(p),i(p+1);i(q),i(q+1)\bigr).
\end{align*}
Here $N_{i(c)}$ denotes the value $N_{i(k)}$ which belongs to the
cycle $c$ of $\gamma\pi$. Note that the appearance of the
$\sigma$-factors ensures that, although $i(k)$ might be different
for different $k$ in the cycle $c$, all $N_{i(k)}$ for $k\in c$ are
actually the same.

Thus in the limit $n\to\infty$, again only the non-crossing pairings
survive and we get as before
$$\lim_{n\to\infty} E[\tr(X_{N_1,\dots,N_d}^m)]=\sum_{\pi\in NC_2(m)} {\mathcal{K}}_\pi,$$
where now the contribution of a $\pi\in NC_2(m)$ is given by
$$
\KK_\pi:=
\sum_{i(1),\dots,i(m)=1}^d\prod_{c\in\gamma\pi}\alpha_{i(c)}
\prod_{(p,q)\in\pi}
 \sigma
\bigl(i(p),i(p+1);i(q),i(q+1)\bigr).$$ As before we can write this as
a trace of a matrix-valued object -- but this time everything is
weighted with the vector $\alpha$. Namely, let us define the matrix
$$\kk_\pi=(\kk_\pi(i,j))_{i,j=1}^d$$
by
\begin{multline*}
\kk_\pi(i,j):= \sum_{i(1)\dots,i(m),i(m+1)=1}^d \delta_{ii(1)}\delta_{ji(m+1)}\\
\prod_{1\not\in c\in\gamma\pi}\alpha_{i(c)} \prod_{(p,q)\in\pi}
 \sigma
\bigl(i(p),i(p+1);i(q),i(q+1)\bigr),
\end{multline*}
where this time we only take the product over those cycles of
$\gamma\pi$ which do not contain the element 1. The factor
$\alpha_{i(1)}$ corresponding to the cycle containing 1 will be used
as a weighting factor for the remaining trace:
$$\KK_\pi=\tr_\alpha (\kk_\pi).$$

One can now check easily that the value of $\kk_\pi$ is calculated
in the same way as before, by an iterated application of the (now
weighted) covariance mapping $\eta_\alpha$, nested according to the
nesting of the blocks of the non-crossing pairing $\pi$.

Let us summarize our calculations: We have
$$\lim_{n\to\infty} E[\tr_{n}(X_{N_1,\dots,N_d}^m)]=
\tr_\alpha\bigl\{ \sum_{\pi\in NC_2(m)} \kk_\pi\bigr\},$$ where
$\kk_\pi$ are $d\times d$ matrices in $M_d^\alpha(\CC)$, determined
in a recursive way as above, and given by an iterated application of
the mapping $\eta_\alpha$.

The derivation of the equation (\ref{eq:rectangular}) is then as
before.

\section{Results and Discussion}\label{sec:results}
Our theorems give us the Cauchy transform $G$ of the asymptotic
eigenvalue distribution of the considered block matrices in the form
$G(z)=\tr_\alpha(\cG(z))$, where $\cG(z)$ is a solution to the
matrix equation (\ref{eq:square}), (\ref{eq:rectangular}), or
(\ref{eq:zwei}). We recover the corresponding eigenvalue
distribution $\mu$ from $G$ in the usual way, by invoking Stieltjes
inversion formula
\begin{equation}
  \label{eq:Stielties Inversion}
  d\mu(x)=-\frac 1\pi \lim_{\ee\searrow 0}\Im G(x+i\ee)dx,
\end{equation} where
the limit is weak convergence of measures (as we will remark in Sec.~\ref{sec:GROC}).

Usually, there is no explicit solution for our matrix equation, so
that we have to rely on numerical methods for solving those. Note
that we do not get directly an equation for $G$. We first have to
solve the matrix equation, then take the (weighted) trace of this
solution. Thus, in terms of the entries of our matrix $\cG$, we
face a system of quadratic equations which we solve numerically using
Newton's algorithm \cite{NumericalRecipes-92}.

In many concrete cases the matrix $\cG$ exhibits some special structure
(i.e., some of its entries might be zero or some of them agree); taking this into
account from the beginning reduces the complexity of the considered system
of equations considerably. This special pattern of the entries of $\cG$ depends
of course on the considered block structure of our problem, i.e. on the covariance
function $\eta$. Let us point out how one can determine this pattern of $\cG$, relying
on the knowledge of $\eta$. Let us concentrate on the equation  \eqref{eq:square} for
square matrices.  As we remarked in the previous section, for large $z$,
one can produce a solution to this equation by iterating the map
$$D\mapsto \frac 1z I_d+\frac 1z \eta(D)\cdot D.$$
Since a pattern in the entries of $\cG$ which is valid for large $z$ must, by analytic
continuation, remain valid everywhere, we can use the above map to determine
this pattern. Since the factor $1/z$ plays no role for recognizing such a pattern, we
start, for example, with the identity matrix $D=I_d$ and iterate the mapping
$D\mapsto I_d +\eta(D)\cdot D$
a few times until the pattern in the entries of $D$ stabilizes. This is then the wanted pattern
for the entries of $\cG$.

In this section we will consider concrete examples of block matrices and compare our solution
from Theorem \ref{thm:square} or \ref{thm:rectangular} with the result of simulations.
First, the well-known Marchenko-Pastur law is derived analytically.
Then the eigenvalue distribution of two different block matrix scenarios
is calculated. We conclude this section with remarks on convergence and the numerics
of the proposed method.

\subsection{Marchenko-Pastur law}
This is a warm-up  example in which we re-derive the celebrated
Marchenko-Pastur law from Theorem \ref{thm:rectangular}.
\subsubsection{Proposition: Marchenko-Pastur}
If $H_N$ is and $M\times N$ matrix of i.i.d. random variables of
unit variance, then as $N\to\infty$, $N/M\to \lambda>0$, the
empirical law of the eigenvalues of $HH^*/M$ converges weakly with
probability one to
\begin{equation}
  \label{eq:Marchenko-Pastur}
  \mu(dx)= (1-\lambda)^+\delta_0+ \frac{\sqrt{(x-1-\lambda)^2-4\lambda}}{2\pi x} 1_{(1-\sqrt{\lambda})^2\leq x\leq (1+\sqrt{\lambda})^2}dx
\end{equation}

\begin{proof} Since this is a well known result, we just show how the final formula follows from
Theorem \ref{thm:rectangular} applied to  $X$ given by \eqref{eq:X}
with  $H$ replaced by $H/\sqrt{M+N}$,   $d=2$,
$\alpha_1=\lim_{M\to\infty}M/(M+N)=1/(1+\lambda)$,
$\alpha_2=\lambda/(1+\lambda)$; we only consider the case
$\lambda\geq 1$. From \eqref{eq:HHst-X2} we have
$$\lim_{N\to\infty}G_{HH^*/(M+N)}(z)=\frac{1+\lambda}{2}\tr_\alpha \cG(z)-\frac{\lambda-1}{2z}=f(z),$$
where $\cG(z)=\mbox{diag}(f(z),g(z))$ satisfies equation
\eqref{eq:zwei} with $\eta(\mbox{diag}(a,b))=\mbox{diag}(\alpha_2
b,\alpha_1 a)$; we  used \eqref{eq:vier} to identify the answer with
$f$. Thus
$$
zf=1+\lambda z fg/(1+\lambda),\; zg=1+ zfg/(1+\lambda).
$$
After a calculation which includes renormalization (dilation by
$1/(1+\lambda)$), this gives
$$\lim_{M\to\infty}G_{HH^*/M}(z)=
\frac{f(z/(1+\lambda))}{1+\lambda}=
\frac{z+(1-\lambda)-\sqrt{(z-1-\lambda)^2-4\lambda}}{2 z},$$ which
is the Cauchy transform of the Marchenko-Pastur law, see for example
\cite[page 102]{Hiai-Petz}
 \end{proof}

\subsection{Block Toeplitz matrices}\label{1Ex} In this
section we show how to use Theorem \ref{thm:square} to identify the
limit laws of block Toeplitz matrices that were considered in \cite{Tamer}.
The interesting feature of this example is that $\cG$ is
non-diagonal. In this example we present only details for the case $d=3$;
in the figures we also show our result and comparison with simulation results
for the cases $d=4$ and $d=5$.

\begin{figure}[hbt]\centering{
\subfigure[$3\times 3$]{\label{fig:3x3}\includegraphics[width=6.5cm]{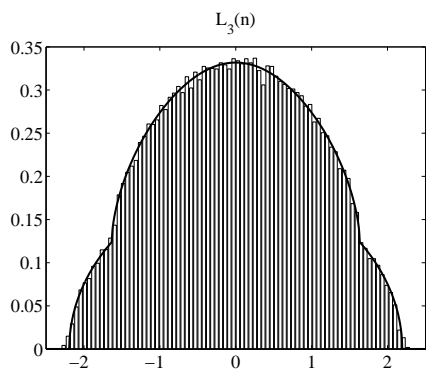}}
\subfigure[$4\times 4$]{\label{fig:4x4}\includegraphics[width=6.5cm]{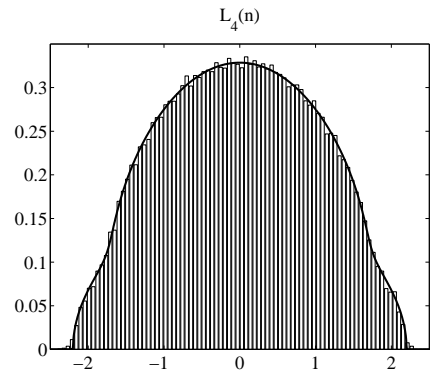} }
\subfigure[$5\times 5$]{\label{fig:5x5}\includegraphics[width=6.5cm]{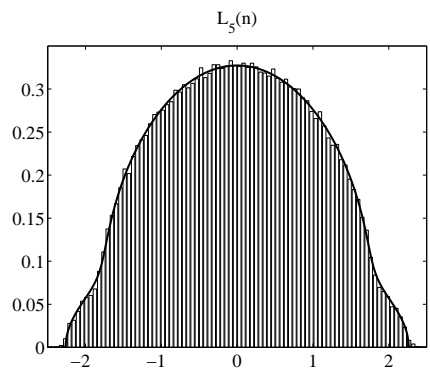} }

}
\caption{\label{Toepl3} Results for block Toeplitz matrices from Section 5.2. The
histogram in each figure stands for the simulation result (100 realization of matrices
with  $100\times 100$ Gaussian blocks) while
the darker curve represents the numerical results of the proposed method.}
\end{figure}

Suppose $A,B,C$ are independent selfadjoint $N\times N$ matrices with i.i.d. entries of unit variance,
and consider the block Toeplitz matrix:
$$X=\frac{1}{\sqrt{3N}}
 \begin{bmatrix} A &B &C \\
B&A&B\\ C&B&A
\end{bmatrix}.$$ Then as $N\to\infty$ the spectral law of $X$ converges almost surely to a deterministic
law, see \cite{Tamer}. Since the limit  does not depend on the
distribution of the entries, we can assume that they are Gaussian
and apply Theorem \ref{thm:square}. This determines the  Cauchy
transform of the limit law as the normalized trace of $\cG$; the latter
is determined by Equation \eqref{eq:square}. For $\cG$ of the form
$$\cG=\begin{bmatrix}
   f & 0 & h  \cr 0 & g & 0  \cr
     h & 0 & f \cr  \end{bmatrix},$$
$\eta$ acts as follows:
 $$ \;\eta(\cG)=\frac13\begin{bmatrix}  2\,f + g & 0 & g + 2\,h \cr 0 & 2\,f +
   g + 2\,h & 0 \cr g + 2\,h & 0 & 2\,f +
   g   \end{bmatrix}.$$
Since the above pattern of $\cG$ is preserved under the mapping $D\mapsto I_3+
\eta(D)\cdot D$, we know that our wanted solution must have this form.
So \eqref{eq:square} gives the following system of equations:
\begin{eqnarray}
    z f &=& 1 + \frac{g\,\left( f + h \right)  +
    2\,\left( f^2 + h^2 \right) }{3},\\
  z g &=& 1 + \frac{g\,
       \left( g + 2\,\left( f + h \right)  \right)
         }{3}, \\ zh  &=&
   \frac{4\,f\,h + g\,\left( f + h \right) }{3}.
\end{eqnarray}
     This system of equations simplifies to two quadratic equations which can be solved by symbolic software;
    the density is a mixture of  the semicircle law and another curve, but the exact solutions
      were not helpful in the selection of the appropriate root.
   Fig.~\ref{fig:3x3} illustrates the histogram and the theoretical
   density obtained numerically from applying the Stieltjes inversion formula on the Cauchy transform
   $G(z)=(2f(z)+g(z))/3$, $\mu\left(t\right)=-\frac 1 \pi \lim_{s\rightarrow 0}\Im G\left(t+is\right)$.
The numerical solution is calculated by solving the resulting system of equations using Newton's method.
The histograms are calculated  for 100 matrices with Gaussian blocks of size $100\times
100$.

In Figures \ref{fig:4x4} and \ref{fig:5x5} we compare the simulation
and the result of our method for the cases $d=4$ with
$$X=\frac 1 {\sqrt{4N}}\begin{bmatrix} A &B &C &D \\
B&A&B&C\\
C&B&A&B\\
D&C&B&A
\end{bmatrix},$$
and $d=5$ with
$$X=\frac 1 {\sqrt{5N}}\begin{bmatrix} A &B &C &D &E\\
B&A&B&C&D\\
C&B&A&B&C\\
D&C&B&A&B\\
E&D&C&B&A
\end{bmatrix}.$$

It might be of interest to point out that the double
limit, first as block sizes $N\to\infty$ and then $d\to\infty$ of
block Toeplitz matrices gives the semicircle law. This can be seen
by passing to $d\times d$ Toeplitz matrices with semicircular
entries. Since the combinatorial estimates from
\cite{Bryc-Dembo-Jiang} apply to this case, the moments are given as
sums over pair partitions. For semicircular elements, only
non-crossing partitions contribute, and each such partitions
contributes $1$.

\subsection{Wishart matrix of rectangular block Toeplitz matrices}
In this section the introduced problem of calculating the spectrum for a MIMO channel is addressed,  where it is of interest to calculate the distribution of the Wishart matrix of rectangular block Toeplitz matrices. First for the case $n_T=n_R$, a proposition is presented that simplifies the process to calculate the desired spectrum. In the Example subsection, first the spectrum for a MIMO system where $n_T=n_R$ is calculated. Then the problem for the $n_R=2n_T$ is tackled and the results are presented.

Assuming $n_T=n_R=N$ and denoting the channel matrix with $H_N$, the following proposition draws the way to calculate the eigenvalue distribution of $H_N^*H_N$.

\subsubsection{Proposition}\label{prop:wireless} As $N\to\infty$,  $d=2K+L-1$, $n=Nd$ the spectral law of
$H_NH_N^*/n$ converges weakly with probability one to the
deterministic probability  measure which is a mixture of $K$
densities with Cauchy-Stieltjes transform
$$
\lim_{N\to\infty}G_{HH^*/n}(z)=\frac{1}{d}\sum_{j=1}^K f_j(z).
$$
Functions $f_j$ are each a Cauchy transform of a probability measure
and the following conditions hold.
\begin{enumerate}
\item $f_j=f_{K+1-j} \mbox{ for } 1\leq j\leq K $
\item There exist Cauchy transforms $g_1,g_2,\dots,g_{K+L-1}$ of some probability measures such that
$g_j=g_{K+L-j} \mbox{ for } 1\leq j \leq K+L-1$ with the property
that
$$\cG(z)=\mbox{diag}(f_1,f_2,\dots,f_K,g_1,g_2,\dots,g_{K+L-1})$$
satisfies \eqref{eq:zwei} with
$$\eta(\cG)_{j,j}=\begin{cases}
\frac1d\sum_{k=0}^{
L-1}g_{j+k}, &\mbox{ if } 1\leq j \leq K,\\
\frac1d\sum_{k=1}^{j-K}f_{k} &\mbox{ if } K+1\leq j \leq K+\min\{K,L\}, \\
\frac1d\sum_{k=j-K-L+1}^{j-K}f_{k} &\mbox{ if } K+L< j \leq 2K\mbox{ and } L< K, \\
\frac1d\sum_{k=1}^{K}f_{k} &\mbox{ if } 2K< j\leq K+L \mbox{ and } L\geq K, \\
\frac1d\sum_{k=j-K-L+1}^{K}f_{k} &\mbox{ if } K+\max\{K,L\}< j \leq
2K+L-1.
\end{cases}$$
\end{enumerate}

\begin{proof} We omit the proof of almost sure convergence, see
Section \ref{sec:GROC}.  To verify that $\cG(z)$ is diagonal and
its diagonal entries satisfy the symmetry condition we compute the
diagonal entries of $\eta(D)$. For $1\leq j \leq K$ we have
$$[\eta(D)]_{j,j}=\frac{1}{d}\sum_{k=
0}^{L-1} d_{K+j+k,K+j+k}.$$ For $1\leq j \leq \min\{K,L\}$ we have
$$[\eta(D)]_{K+j,K+j}=\frac{1}{d}\sum_{k=1}^j d_{k,k}.$$
For $\min\{K,L\}\leq j \leq \max\{K,L\}$  we have
$$[\eta(D)]_{K+j,K+j}=
\begin{cases}
\frac{1}{d}\sum_{k=j-L+1}^{j} d_{k,k}& L<K\\
 \frac{1}{d}\sum_{k=1}^{K} d_{k,k} &L\geq K. \end{cases}$$
For $\max\{K,L\}\leq j \leq K+L-1$  we have
$$[\eta(D)]_{K+j,K+j}=\frac{1}{d}\sum_{k=j-L+1 }^{K} d_{k,k}.$$
We note that the symmetry conditions $f_j=f_{K+1-j}$ and
$g_j=g_{K+L-j}$ are
preserved under the mapping $D\mapsto I_d+\eta(D)\cdot D$. Since $\eta$ maps
diagonal matrices into
diagonal matrices, our solution $\cG$ must be diagonal and the diagonal
entries must satisfy the symmetry conditions as claimed.
\end{proof}
\subsubsection{Examples}
\begin{figure}[hbt]\centering{
\includegraphics[width=8cm]{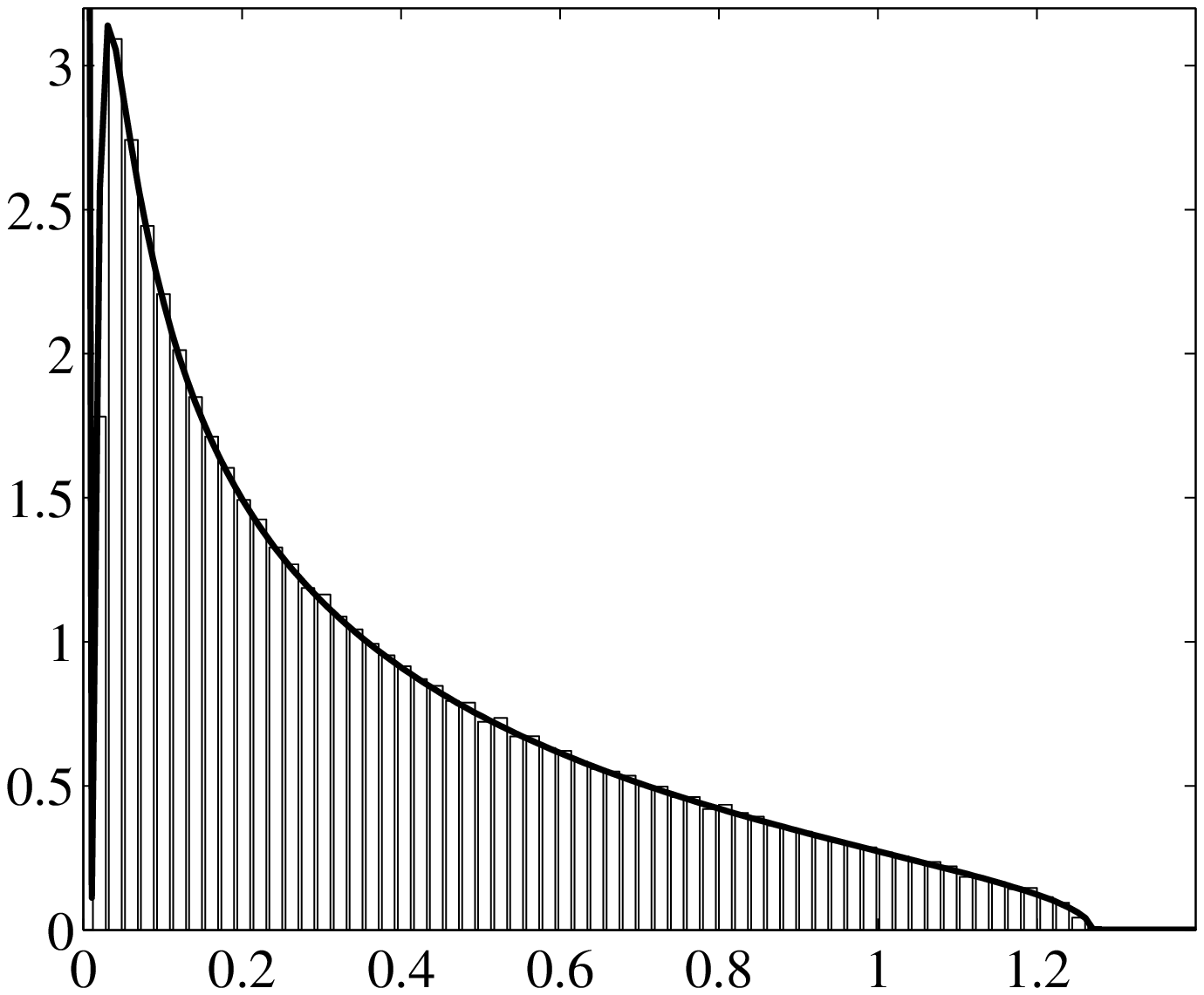}}
\caption{\label{ABCD000}Superimposed theoretical density of the eigenvalues
of complex normal $H_nH_n^*/n$ for a channel with ISI $L=4$ and a MIMO system $n_R=n_T$ with frame length of $K=4$ over its histogram for $N=100$, based on $100$ realizations.}
\end{figure}

In this part, first the developed theorems are used toward calculation of spectrum of a MIMO system with ISI ($L=4$) and frame size of 4 ($K=4$) while $n_T=n_R$. Then, for the simpler case of $K=2$ and $L=2$ when $n_R=2n_T$ (or $n_T=2n_R$) the spectrum is calculated.
\newline Consider
$$
H_N=\left[\begin{array}{*{20}c}
   A & B & C & D & 0 & 0 & 0 \\ 0 & A & B & C &
   D & 0 & 0 \\ 0 & 0 & A & B & C & D & 0 \\ 0 &
   0 & 0 & A & B & C & D \\
   \end{array}\right],
$$
where $A,B,C,D$ are independent non-selfadjoint Gaussian $N\times
N$-random matrices. It is also assumed that the impulse response of the channel from any transmitter antenna to any receiver antenna is identical and equal to $\left[\begin{array}{cccc}1& 1& 1& 1\end{array}\right]$. In this case $K=L=4$, $d=11$,
$$\cG(z)=\mbox{diag}(f_1,f_2,f_2,f_1,g_1,g_2,g_3,g_4,g_3,g_2,g_1),$$
and \eqref{eq:zwei} yields the following system of equations
\begin{eqnarray}
zf_1&=&1 + zf_1 (g_1 + g_2 + g_3 + g_4)/11 \label{eq:f1}
\\
zf_2&=&1 + zf_2 (g_2 + 2 g_3 + g_4)/11
\\
zg_1&=&1 + zf_1 g_1/11
\\zg_2&=&1 + z(f_1 + f_2) g_2/11
\\zg_3&=&1 + z(f_1 + 2 f_2) g_3/11
\\
zg_4&=&1 + 2 z(f_1 + f_2) g_4/11.\label{eq:g4}
\end{eqnarray}
The limiting Cauchy transform is $G_{HH^*}(z)=(f_1+f_2)/2$. After
eliminating $g_j$ from the equations we get
\begin{eqnarray*}
z &=& \frac1{f_1} + \frac{1}{11 - f_1}+\frac{1}{11 - f_1 - f_2}
+\frac{1}{11 - f_1 - 2f_2} + \frac{1}{11 - 2f_1 - 2f_2},\\
 z &=&
\frac1{f_2} + \frac1{11 - f_1 - f_2} +\frac2{11 - f_1 - 2f_2} +
\frac1{11 - 2f_1 - 2f_2}.
\end{eqnarray*}
Mathematica failed to solve this system in a reasonable time. Fortunately, the
Newton's algorithm can be applied to the equations \eqref{eq:f1}-\eqref{eq:g4} and its answer matches the
simulations (Fig.~\ref{ABCD000}).

\begin{figure}[hbt]\centering{
\includegraphics[width=8cm]{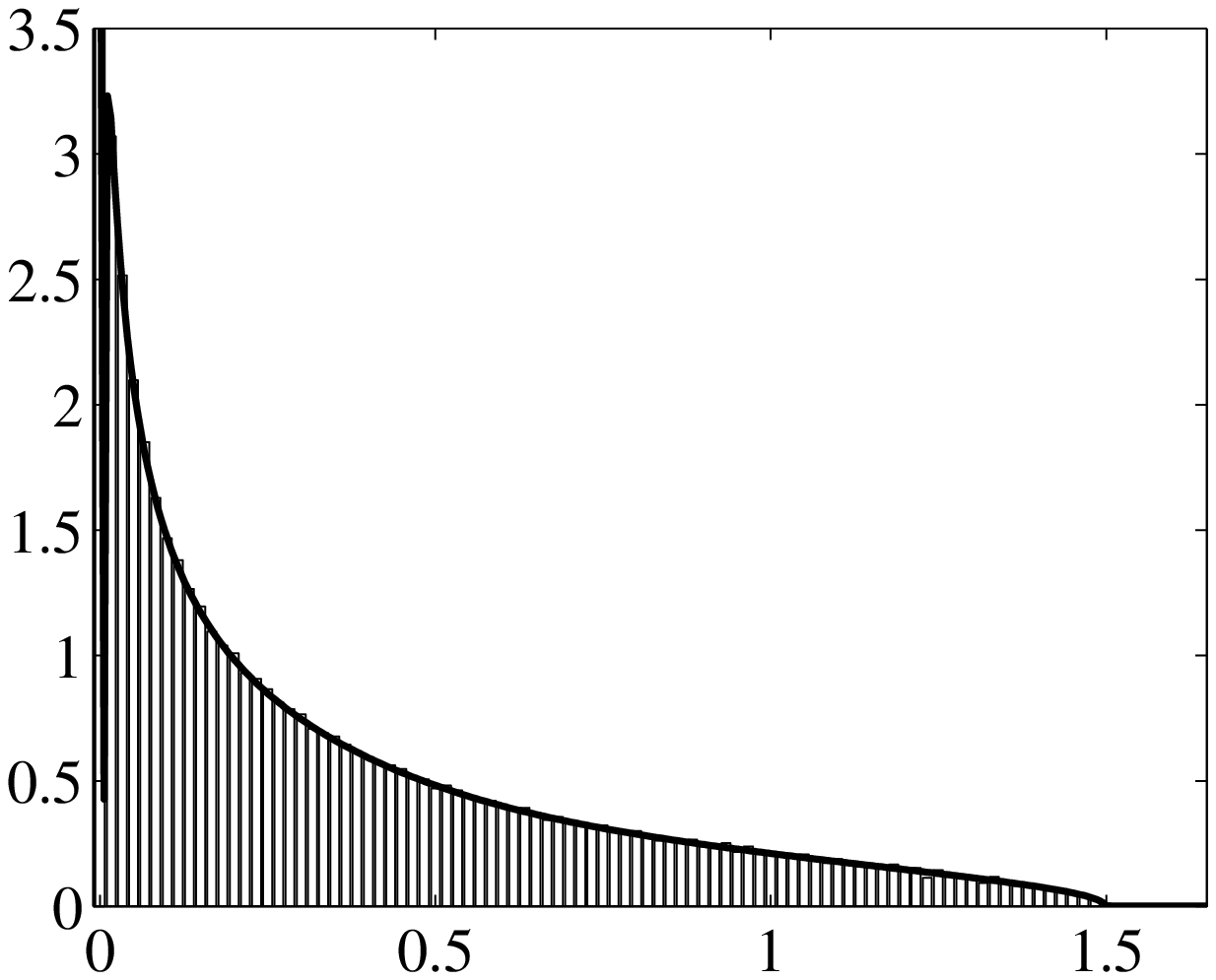}}
\caption{\label{AB0_CD0}Superimposed theoretical density of the eigenvalues
of complex normal $H_nH_n^*/n$ for a channel with ISI $L=2$ and a MIMO system $n_R=2n_T$ with frame length of $K=2$ over its histogram for $N=100$, based on $100$ realizations. }
\end{figure}

On the other hand, the channel matrix for a MIMO system with $n_R=2n_T$ and frame length of $K=2$ over a channel with ISI ($L=2$) is as follows:
$$H_N=\left[\begin{array}{ccc}
A& B&0\\
0&A&B\end{array}\right],$$
where $A$ and $B$ are independent non-selfadjoint $2N\times N$ Gaussian random matrices. While one may calculate the spectrum of $HH^*$ using Theorem~\ref{thm:rectangular}, it is also possible to consider \begin{eqnarray}\label{double_ratio}H_N=\left[\begin{array}{ccc}
A& B&0\\C&D&0\\
0&A&B\\0&C&D\end{array}\right],\end{eqnarray}
for $A$, $B$, $C$ and $D$ independent non-selfadjoint $N\times N$ Gaussian random matrices and calculate the spectrum of $HH^*$ using Theorem~\ref{thm:square}. The spectrum of $HH^*$ is calculated using the Newton's method which is depicted in Fig.~\ref{AB0_CD0}.

\subsection{General remarks on convergence}\label{sec:GROC}
 In Theorems \ref{thm:square} and
\ref{thm:rectangular}, the convergence means weak convergence of the
expected values of the empirical laws of the eigenvalues, which is
sufficient to identify the limit. In applications,  the appropriate
mode of convergence is weak convergence of random empirical measures
with probability one. Such convergence means that for a "typical"
realization of a random matrix, the limiting law as determined from
Theorems \ref{thm:square} and \ref{thm:rectangular} is a good
approximation when the size of the random matrix is large enough.

Results of this type are readily available for symmetric independent
blocks with i.i.d. entries, see \cite{Tamer}.  In the Gaussian case
one can easily extend such results to non-Hermitian and dependent
square blocks by noting the following:
\begin{itemize}
\item If $A$ consists of  i.i.d. Gaussian random variables, then
$A=A'/\sqrt{2}+A''/(\sqrt{2}i)$ where $A'=(A+A^*)/\sqrt{2}$ and
$A''=i(A-A^*)/\sqrt{2}$ are independent, Hermitian, and their
off-diagonal entries are i.i.d Gaussian with the same variance as
the entries of $A$.
\item If $A,B$ are jointly Gaussian with respective entries that share the same
correlation $\sigma$ (and are otherwise i.i.d), then
$B=A+\sqrt{1-\sigma^2}B'$, where $B'=(A-\sigma B)/\sqrt{1-\sigma^2}$
is independent of $A$ and has i.i.d. entries of the same variance as
the entries of $A$.
\end{itemize}
For a Hermitian block matrix $X$ consisting of square blocks, these
two transformations allow us to express $\tr_n(X^k)$ as a linear
combination of the polynomials in independent Gaussian and Hermitian
matrices. Under appropriate normalization (i.e. when the matrices
are constructed from an infinite i.i.d family of unit variance, and
the matrices are scaled by the square root of the dimension), the
latter converge almost surely, see for example \cite{Hiai-Petz00a}.
Since the limiting law is compactly supported, weak convergence of
empirical measures holds with probability one.

\subsection{General remarks on the numerics}\label{sec:GRON}
The main inherent problem is that we have to find the "right" root.
As we do not have a general automated recipe for doing this, we
choose our solution by inspection.
 Theorem \ref{thm:square} guarantees uniqueness of an
analytic solution under condition \eqref{eq:cG lim}, but the
validity of this condition is difficult to judge for a numerical
solutions. It seems that in general there is only one solution which
corresponds to a valid probability measure. Furthermore, as it is quite
easily seen, the fact that $\cG$ is essentially a generating power
series in operator-valued moments implies that each diagonal entry
of $\cG$ is itself the Cauchy transform of some probability measure.
Thus the solution we are looking for must exhibit the very strong
property that all diagonal entries of $\cG(z)$ yield a valid
probability distribution when $z$ approaches the real line.

Our experience with this is reflected quite adequately in the
following remark of Edelman and Rao \cite{Edelman-Rao-06} on the
same kind of problem with their random matrix calculator:
``Irrespective of whether the encoded probability measure is
compactly supported or not, the $-1/z$ behaviour of the real part of
Stieltjes transform (the principal value) as $z\to\pm\infty$ helps
identify the correct root. In our experience, while multiple root
curves might exhibit this behavior, invariably only one root will
have an imaginary branch that, when normalized, will correspond to a
valid probability measure. Why this always appears to be the case
for the operational laws described is a bit of a mystery to us."

\section{Conclusion}\label{sec:conclusion}
We use operator-valued free probability theory and develop a
method to calculate the limiting spectra for block matrices.
Using this method we
compute the eigenvalue distribution for
block-matrices arising in wireless communications. The
agreement with simulation results is excellent. We expect our method
to have a much wider applicability.

\subsection*{Acknowledgement}
We thank Ralf M\"uller for bringing the problem of block matrices in
the context of MIMO to our attention.

\clearpage

\clearpage

\listoffigures
\end{document}

%% file: bilder.tex
\setlength{\unitlength}{0.4cm}

\newsavebox{\Catze}
\savebox{\Catze}(2,2){
 \thinlines
\multiput(0,0)(1,0){3}{\line(0,1){2}}
\multiput(0,0)(0,1){3}{\line(1,0){2}} \thicklines
\put(0,0){\vector(1,1){1}} \put(1,1){\vector(1,-1){1}}}

\newsavebox{\NCze}
\savebox{\NCze}(2,2){
\thicklines
\put(0,0){\line(1,0){1}}
\put(0,0){\line(0,1){1}}
\put(1,0){\line(0,1){1}}}

\newsavebox{\Catve}
\savebox{\Catve}(4,3){
 \thinlines
\multiput(0,0)(1,0){5}{\line(0,1){3}}
\multiput(0,0)(0,1){4}{\line(1,0){4}} \thicklines
\put(0,0){\vector(1,1){1}} \put(1,1){\vector(1,-1){1}}
\put(2,0){\vector(1,1){1}} \put(3,1){\vector(1,-1){1}}}

\newsavebox{\Catvz}
\savebox{\Catvz}(4,3){
 \thinlines
\multiput(0,0)(1,0){5}{\line(0,1){3}}
\multiput(0,0)(0,1){4}{\line(1,0){4}} \thicklines
\put(0,0){\vector(1,1){1}} \put(1,1){\vector(1,1){1}}
\put(2,2){\vector(1,-1){1}} \put(3,1){\vector(1,-1){1}}}

\newsavebox{\NCvz}
\savebox{\NCvz}(3,3){
\thicklines
\put(0,0){\line(0,1){2}}
\put(0,0){\line(1,0){3}}
\put(3,0){\line(0,1){2}}
\put(1,1){\line(0,1){1}}
\put(1,1){\line(1,0){1}}
\put(2,1){\line(0,1){1}}}

\newsavebox{\NCve}
\savebox{\NCve}(3,3){
\thicklines
\put(0,0){\line(0,1){1}}
\put(0,0){\line(1,0){1}}
\put(1,0){\line(0,1){1}}
\put(2,0){\line(0,1){1}}
\put(2,0){\line(1,0){1}}
\put(3,0){\line(0,1){1}}}

\newsavebox{\Catsf}
\savebox{\Catsf}(6,4){
 \thinlines
\multiput(0,0)(1,0){7}{\line(0,1){4}}
\multiput(0,0)(0,1){5}{\line(1,0){6}} \thicklines
\put(0,0){\vector(1,1){1}} \put(1,1){\vector(1,1){1}}
\put(2,2){\vector(1,1){1}} \put(3,3){\vector(1,-1){1}}
\put(4,2){\vector(1,-1){1}} \put(5,1){\vector(1,-1){1}}}

\newsavebox{\Catsv}
\savebox{\Catsv}(6,4){
 \thinlines
\multiput(0,0)(1,0){7}{\line(0,1){4}}
\multiput(0,0)(0,1){5}{\line(1,0){6}} \thicklines
\put(0,0){\vector(1,1){1}} \put(1,1){\vector(1,1){1}}
\put(2,2){\vector(1,-1){1}} \put(3,1){\vector(1,1){1}}
\put(4,2){\vector(1,-1){1}} \put(5,1){\vector(1,-1){1}}}

\newsavebox{\Catsd}
\savebox{\Catsd}(6,4){
 \thinlines
\multiput(0,0)(1,0){7}{\line(0,1){4}}
\multiput(0,0)(0,1){5}{\line(1,0){6}} \thicklines \thicklines
\put(0,0){\vector(1,1){1}} \put(1,1){\vector(1,1){1}}
\put(2,2){\vector(1,-1){1}} \put(3,1){\vector(1,-1){1}}
\put(4,0){\vector(1,1){1}} \put(5,1){\vector(1,-1){1}}}

\newsavebox{\Catsz}
\savebox{\Catsz}(6,4){
 \thinlines
\multiput(0,0)(1,0){7}{\line(0,1){4}}
\multiput(0,0)(0,1){5}{\line(1,0){6}} \thicklines
\put(0,0){\vector(1,1){1}} \put(1,1){\vector(1,-1){1}}
\put(2,0){\vector(1,1){1}} \put(3,1){\vector(1,1){1}}
\put(4,2){\vector(1,-1){1}} \put(5,1){\vector(1,-1){1}}}

\newsavebox{\Catse}
\savebox{\Catse}(6,4){
 \thinlines
\multiput(0,0)(1,0){7}{\line(0,1){4}}
\multiput(0,0)(0,1){5}{\line(1,0){6}}  \thicklines
\put(0,0){\vector(1,1){1}} \put(1,1){\vector(1,-1){1}}
\put(2,0){\vector(1,1){1}} \put(3,1){\vector(1,-1){1}}
\put(4,0){\vector(1,1){1}} \put(5,1){\vector(1,-1){1}}}

\newsavebox{\NCsf}
\savebox{\NCsf}(6,4){
\thicklines
\put(0,0){\line(0,1){3}}
\put(0,0){\line(1,0){5}}
\put(5,0){\line(0,1){3}}
\put(1,1){\line(0,1){2}}
\put(1,1){\line(1,0){3}}
\put(4,1){\line(0,1){2}}
\put(2,2){\line(1,0){1}}
\put(2,2){\line(0,1){1}}
\put(3,2){\line(0,1){1}}}

\newsavebox{\NCsv}
\savebox{\NCsv}(6,4){
\thicklines
\put(0,0){\line(0,1){2}}
\put(0,0){\line(1,0){5}}
\put(5,0){\line(0,1){2}}
\put(1,1){\line(0,1){1}}
\put(1,1){\line(1,0){1}}
\put(2,1){\line(0,1){1}}
\put(3,1){\line(1,0){1}}
\put(3,1){\line(0,1){1}}
\put(4,1){\line(0,1){1}}}

\newsavebox{\NCsd}
\savebox{\NCsd}(6,4){
\thicklines
\put(0,0){\line(0,1){2}}
\put(0,0){\line(1,0){3}}
\put(3,0){\line(0,1){2}}
\put(1,1){\line(0,1){1}}
\put(1,1){\line(1,0){1}}
\put(2,1){\line(0,1){1}}
\put(4,0){\line(0,1){2}}
\put(4,0){\line(1,0){1}}
\put(5,0){\line(0,1){2}}}

\newsavebox{\NCsz}
\savebox{\NCsz}(6,4){
\thicklines
\put(0,0){\line(0,1){2}}
\put(0,0){\line(1,0){1}}
\put(1,0){\line(0,1){2}}
\put(2,0){\line(0,1){2}}
\put(2,0){\line(1,0){3}}
\put(5,0){\line(0,1){2}}
\put(3,1){\line(0,1){1}}
\put(3,1){\line(1,0){1}}
\put(4,1){\line(0,1){1}}}

\newsavebox{\NCse}
\savebox{\NCse}(6,4){
\thicklines
\put(0,0){\line(0,1){1}}
\put(0,0){\line(1,0){1}}
\put(1,0){\line(0,1){1}}
\put(2,0){\line(0,1){1}}
\put(2,0){\line(1,0){1}}
\put(3,0){\line(0,1){1}}
\put(4,0){\line(0,1){1}}
\put(4,0){\line(1,0){1}}
\put(5,0){\line(0,1){1}}}

%% file: bild2.tex
\setlength{\unitlength}{0.4cm}

\newsavebox{\NCzeZ}
\savebox{\NCzeZ}(2,2){
\thicklines
\put(0,0){\line(1,0){1}}
\put(0,0){\line(0,1){1}}
\put(1,0){\line(0,1){1}}
\put(0,1.7){\makebox(0,0){1}}
\put(1,1.7){\makebox(0,0){2}}}

\newsavebox{\NCvzZ}
\savebox{\NCvzZ}(3,3){
\thicklines
\put(0,0){\line(0,1){2}}
\put(0,0){\line(1,0){3}}
\put(3,0){\line(0,1){2}}
\put(1,1){\line(0,1){1}}
\put(1,1){\line(1,0){1}}
\put(2,1){\line(0,1){1}}
\put(0,2.7){\makebox(0,0){1}}
\put(1,2.7){\makebox(0,0){2}}
\put(2,2.7){\makebox(0,0){3}}
\put(3,2.7){\makebox(0,0){4}}}

\newsavebox{\NCveZ}
\savebox{\NCveZ}(3,3){
\thicklines
\put(0,0){\line(0,1){1}}
\put(0,0){\line(1,0){1}}
\put(1,0){\line(0,1){1}}
\put(2,0){\line(0,1){1}}
\put(2,0){\line(1,0){1}}
\put(3,0){\line(0,1){1}}
\put(0,1.7){\makebox(0,0){1}}
\put(1,1.7){\makebox(0,0){2}}
\put(2,1.7){\makebox(0,0){3}}
\put(3,1.7){\makebox(0,0){4}}}

\newsavebox{\NCsfZ}
\savebox{\NCsfZ}(6,4){
\thicklines
\put(0,0){\line(0,1){3}}
\put(0,0){\line(1,0){5}}
\put(5,0){\line(0,1){3}}
\put(1,1){\line(0,1){2}}
\put(1,1){\line(1,0){3}}
\put(4,1){\line(0,1){2}}
\put(2,2){\line(1,0){1}}
\put(2,2){\line(0,1){1}}
\put(3,2){\line(0,1){1}}
\put(0,3.7){\makebox(0,0){1}}
\put(1,3.7){\makebox(0,0){2}}
\put(2,3.7){\makebox(0,0){3}}
\put(3,3.7){\makebox(0,0){4}}
\put(4,3.7){\makebox(0,0){5}}
\put(5,3.7){\makebox(0,0){6}}}

\newsavebox{\NCsvZ}
\savebox{\NCsvZ}(6,4){
\thicklines
\put(0,0){\line(0,1){2}}
\put(0,0){\line(1,0){5}}
\put(5,0){\line(0,1){2}}
\put(1,1){\line(0,1){1}}
\put(1,1){\line(1,0){1}}
\put(2,1){\line(0,1){1}}
\put(3,1){\line(1,0){1}}
\put(3,1){\line(0,1){1}}
\put(4,1){\line(0,1){1}}
\put(0,2.7){\makebox(0,0){1}}
\put(1,2.7){\makebox(0,0){2}}
\put(2,2.7){\makebox(0,0){3}}
\put(3,2.7){\makebox(0,0){4}}
\put(4,2.7){\makebox(0,0){5}}
\put(5,2.7){\makebox(0,0){6}}}

\newsavebox{\NCsdZ}
\savebox{\NCsdZ}(6,4){
\thicklines
\put(0,0){\line(0,1){2}}
\put(0,0){\line(1,0){3}}
\put(3,0){\line(0,1){2}}
\put(1,1){\line(0,1){1}}
\put(1,1){\line(1,0){1}}
\put(2,1){\line(0,1){1}}
\put(4,0){\line(0,1){2}}
\put(4,0){\line(1,0){1}}
\put(5,0){\line(0,1){2}}
\put(0,2.7){\makebox(0,0){1}}
\put(1,2.7){\makebox(0,0){2}}
\put(2,2.7){\makebox(0,0){3}}
\put(3,2.7){\makebox(0,0){4}}
\put(4,2.7){\makebox(0,0){5}}
\put(5,2.7){\makebox(0,0){6}}}

\newsavebox{\NCszZ}
\savebox{\NCszZ}(6,4){
\thicklines
\put(0,0){\line(0,1){2}}
\put(0,0){\line(1,0){1}}
\put(1,0){\line(0,1){2}}
\put(2,0){\line(0,1){2}}
\put(2,0){\line(1,0){3}}
\put(5,0){\line(0,1){2}}
\put(3,1){\line(0,1){1}}
\put(3,1){\line(1,0){1}}
\put(4,1){\line(0,1){1}}
\put(0,2.7){\makebox(0,0){1}}
\put(1,2.7){\makebox(0,0){2}}
\put(2,2.7){\makebox(0,0){3}}
\put(3,2.7){\makebox(0,0){4}}
\put(4,2.7){\makebox(0,0){5}}
\put(5,2.7){\makebox(0,0){6}}}

\newsavebox{\NCseZ}
\savebox{\NCseZ}(6,4){
\thicklines
\put(0,0){\line(0,1){1}}
\put(0,0){\line(1,0){1}}
\put(1,0){\line(0,1){1}}
\put(2,0){\line(0,1){1}}
\put(2,0){\line(1,0){1}}
\put(3,0){\line(0,1){1}}
\put(4,0){\line(0,1){1}}
\put(4,0){\line(1,0){1}}
\put(5,0){\line(0,1){1}}
\put(0,1.7){\makebox(0,0){1}}
\put(1,1.7){\makebox(0,0){2}}
\put(2,1.7){\makebox(0,0){3}}
\put(3,1.7){\makebox(0,0){4}}
\put(4,1.7){\makebox(0,0){5}}
\put(5,1.7){\makebox(0,0){6}}}